\documentclass[useAMS,usenatbib,usegraphicx]{mn2e}
\usepackage{times}
\usepackage{subfigure}
\usepackage{float}
\usepackage{graphicx}
\usepackage{pdfpages}
\usepackage{adjustbox}
\usepackage{amsmath}
\usepackage{amssymb}
\usepackage{hyperref}
\usepackage{booktabs}
\usepackage{makecell}
\usepackage{lipsum}

\newcommand{\ms}{M$_{\odot}$}
\newcommand{\rs}{R$_{\odot}$}

\title{Properties of the post in-spiral common envelope ejecta II: dust formation}
\author[Iaconi et al.]{Roberto Iaconi $^{1}$\thanks{email: roberto.iaconi@kusastro.kyoto-u.ac.jp} \thanks{JSPS International Research Fellow (Graduate School of Science, Kyoto University)}, Keiichi Maeda$^{1}$, Takaya Nozawa $^{2}$, Orsola De Marco $^{3,4}$ \newauthor
and Thomas Reichardt $^{3,4}$\\
$^{1}$Department of Astronomy, Kyoto University, Kitashirakawa-Oiwake-cho, Sakyo-ku, Kyoto 606-8502, Japan 0000-0002-1940-1950\\
$^{2}$Division of Theoretical Astronomy, National Astronomical Observatory of Japan, Mitaka, Tokyo 181-8588, Japan\\
$^{3}$Department of Physics \& Astronomy, Macquarie University, Sydney, NSW 2109, Australia\\
$^{4}$Astronomy, Astrophysics and Astrophotonics Research Centre, Macquarie University, Sydney, NSW 2109, Australia\\}

\begin{document}

\date{Accepted by MNRAS, \today}

\pagerange{\pageref{firstpage}--\pageref{lastpage}} \pubyear{\the\year{}}

\maketitle

\label{firstpage}

\begin{abstract}
We study the formation of dust in the expanding gas ejected as a result of a common envelope binary interaction. In our novel approach, we apply the dust formation model of Nozawa et al. to the outputs of the 3D hydrodynamic SPH simulation performed by Iaconi et al., that involves a giant of 0.88~\ms \ and 83~\rs, with a companion of 0.6~\ms \ placed on the surface of the giant in circular orbit. After simulating the dynamic in-spiral phase we follow the expansion of the ejecta for $\simeq 18\,000$~days. During this period the gas is able to cool down enough to reach dust formation temperatures.
Our results show that dust forms efficiently in the window between $\simeq 300$~days (the end of the dynamic in-spiral) and $\simeq 5000$~days. The dust forms in two separate populations; an outer one in the material ejected during the first few orbits of the companion inside the primary's envelope and an inner one in the rest of the ejected material. 
We are able to fit the grain size distribution at the end of the simulation with a double power law. The slope of the power law for smaller grains is flatter than that for larger grains, creating a knee-shaped distribution. The power law indexes are however different from the classical values determined for the interstellar medium.
We also estimate that the contribution to cosmic dust by common envelope events is not negligible and comparable to that of novae and supernovae.
\end{abstract}



\section{Introduction}
\label{sec:introduction}
The common envelope interaction (\citealt{Paczynski1976}, \citealt{Ivanova2013}; hereafter CE) is a  binary interaction process that leads to a reduction in a binary's orbital separation resulting in a merger or in the formation of compact evolved binaries, which in turn can becomes Type Ia supernovae, gamma-ray bursts, merging double black holes/neutron stars emitting detectable gravitational waves. A typical common envelope  configuration is a giant primary and a more compact companion, such as a main sequence star or a white dwarf but other configurations are possible. 

Given its intrinsic three-dimensional structure and physical complexity, research on CE has been mainly carried out through hydrodynamic simulations. Several simulations have been performed with different codes (e.g., \citealt{Sandquist1998}, \citealt{Ricker2012}, \citealt{Passy2012} and \citealt{Ohlmann2016}, to cite a few; see table~A3 of \citealt{Iaconi2019b} for a more detailed list). Different physical mechanisms and their impact on the final outcome have been considered: H and He recombination energy injected in the gas when the envelope expands and cools down (\citealt{Nandez2015}, \citealt{Nandez2016} and \citealt{Ivanova2016}, \citealt{Ivanova2018} support the idea that the energy released by recombination has a major contribution in the envelope unbinding process, while \citealt{Grichener2018} and \citealt{Soker2018} support the idea that such energy is mostly radiated away and does not affect strongly the envelope unbinding process); the magnetic field generated by the companion star (\citealt{Ohlmann2016b}); jets from the companion star (\citealt{Shiber2017}, \citealt{Shiber2018}, \citealt{Shiber2019}, \citealt{Schreier2019}, \citealt{Lopez-camara2019}); pre-CE Roche lobe overflow (\citealt{Reichardt2019}); envelope fall-back (\citealt{Kuruwita2016}); high orbital eccentricities (\citealt{Staff2016a}); dust-driven winds (e.g., \citealt{Glanz2018}); stellar pulsations (e.g., \citealt{Clayton2017}); convection (\citealt{Wilson2019}). CE wind tunnel simulations have also been carried out by \citet{MacLeod2017b}, \citet{Murguia-Berthier2017} and \citet{De2019}.
All of these work analyse the dynamic in-spiral phase, during which the two stars quickly approach each other in a time-scale comparable to the dynamical time-scale of the primary, i.e., from months to years (but see, e.g., \citealt{Soker1993} and \citealt{Soker2017} for studies on the post dynamic in-spiral phase).

This paper follows on from \citet{Iaconi2019}, where we analysed the behaviour of the CE ejecta after the dynamic in-spiral. We carried out a 3D SPH simulation almost identical to those of \citet{Iaconi2018}, but we ran the simulation for longer, $18\,434$~days ($\simeq 50$~yr) after the end of the dynamic in-spiral and analysed the dynamic and thermodynamic properties of the extended envelope gas. In that paper we showed that the evolution of the CE ejecta, after the termination of the dynamic in-spiral, can be approximated by homologous expansion
after $\simeq 5000$~days from the beginning of the simulation (the end of in-spiral is at $\simeq 300$~days). We also observed the formation of ring-like features expanding self-similarly. 
If the post dynamic in-spiral CE evolution is homologous, then it can be fully calculated analytically by applying the homologous expansion equations to any simulation data dump chosen as initial conditions. We showed the power of this approach by building a toy model for the calculation of the photosphere where the ejecta expand following the homologous dynamics and the main contribution to opacity is provided by dust formed in the cooling envelope gas. The determination of the photosphere location and temperature carried in \citet{Iaconi2019} did not explain in detail how the calculations of the dust formation were performed, something that we concentrate on here. 

The number of common envelope events observed is multiplying quickly, with most of the observational counterparts of CE events being associated with red nova outbursts (\citealt{Ivanova2013b}). Moreover, a new class of infrared transients has been discovered and classified by \citet{Jencson2019} which show similarities to other events that, in the optical, have been interpreted as common envelope mergers. 
It is also clear that in both of these types of transients dust is often present, with many red novae being fully or partially embedded in dust (e.g.: V4332 Sgr, \citealt{Banerjee2007,Banerjee2015}; BLG-360, \citealt{Tylenda2013}, Kaminski - private communication; NN Ser, \citealt{Hardy2016}; V1309 Sco, \citealt{Tylenda2016}) and several of the new transients showing distinctive dust signatures (\citealt{Jencson2019,Pastorello2019}). 
Therefore, it is essential to study the formation and evolution of dust in the CE environment to fully understand the evolution of these observational counterparts.

Another reason to include dust in CE interactions is that a dusty common envelope may be ejected more readily by radiation pressure. The question of what forces play a role in ejecting the CE is still open \citep[e.g.,][]{Reichardt2020} and dust-driving remains a candidate at least for some of the cases. While in this paper we do not calculate the radiative forces, determining the location, timing and type of dust that forms is a necessary calculation.

A model for dust formation in CE ejecta was previously carried out by \citet{Lu2013}. Their model presents some substantial differences with respect to ours. First, they assume that the CE ejecta expand following the kinematics of a stellar wind with spherical symmetry. Second, they considered olivine-type silicates, pyroxene-type silicates and iron grains, while we utilise the pyroxene-type silicate MgSiO$_3$ and carbon grains. Finally, \citet{Lu2013} adopted the dust formation model by \citet{Ferrarotti2006}, who specify the number of seed nuclei by hand, while we use the dust formation model by \citet{Nozawa2013}, which treats seed formation and dust growth self-consistently.
Dust formation in common envelopes was also considered analytically by \citet{Soker1998} and \citet{Glanz2018}. \citet{Soker1998} proposes the formation of dust taking place in cold spots of the envelope ejecta, following the CE between an AGB star and a Jupiter size object. \citet{Glanz2018} explore instead the idea that dust-driven winds taking place after the dynamic in-spiral might help to unbind the envelope.
Our study, on the other hand, is the first that presents a detailed model of dust formation in CE ejecta based on the outputs of 3D SPH hydrodynamic simulations (we also highlight that, even in other astrophysical contexts, e.g., supernovae, only a few studies of dust formation applied to 3D simulations have been performed). Dust formation does not happen at fixed locations but is achieved in the flow of the gas parcels. Therefore, grid-based hydrodynamic codes, which use Eulerian schemes, can be coupled to dust formation models only by introducing a large number of test particles. Such approach results in approximations due to the fact that no matter how many test particles one introduces, they will not be able to track accurately all the locations of the gas parcels.  On the other hand, this study uses an SPH code, based on a Lagrangian formalism, which allows us to follow every moving element of the fluid without any approximation.

This paper is structured as follows: in Section~\ref{sec:dust_model} we describe the dust formation model we use, how we applied it to our data and what are the effects our numerical choices on it; in Section~\ref{sec:dust_formation_process} we describe the process of dust formation; in Section~\ref{sec:characteristics_of_the_dust} we show the main characteristics of the dust grains that form in the ejected envelope (grains size, location, mass, size distribution and contribution to the cosmic dust); in Section~\ref{sec:observational_counterparts} we compare our results with the observations of dust in the post-CE system V1309 Sco. Finally, the summary and conclusions are given in Section~\ref{sec:summary_and_conclusions}.


\section{Dust formation model}
\label{sec:dust_model}
In this section we give an explanation of the dust formation model and how we applied it to our data. For a full description of the model, see \citet{Nozawa2013}.

\subsection{Oxygen and Carbon-based dust chemistries}
\label{ssec:oxygen_and_carbon_based_dust_chemistries}
\citet{Nozawa2013} formulated the non-steady-state dust formation process involving nucleation and grain growth, and applied it to two grain species: pyroxene-type silicates (MgSiO$_3$) and carbon (C). The choice is based on the fact that these are considered to be the most representative species of dust grains in interstellar space. 
Whether C or MgSiO$_3$ grains are formed is determined by the number ratio of carbon and oxygen atoms in the gas phase (hereafter referred to as the C/O ratio).
The abundances of gaseous C and O atoms are mainly controlled by the formation of CO molecules, because CO molecules efficiently form at relatively high temperatures and can trap almost completely either C or O atoms prior to dust formation, depending on which of the two has a lower number density. Let us consider the conditions required for the formation of the two types of dust separately.

MgSiO$_3$ grains are the most expected type of dust in oxygen-rich envelopes of RGB CE we simulate here, because these stars have not yet undergone the third dredge up and have solar abundances (\citealt{Asplund2009}; where the number fractions of the elements forming the grains considered are ordered in the following way: $n_{\mathrm{O}} > n_{\mathrm{C}} > n_{\mathrm{Mg}} > n_{\mathrm{Si}}$). 
Thus, the fact the C/O ratio is smaller than unit, results in all the C atoms to be locked up in CO molecules. This prevents the formation of C grains, leaving the MgSiO$_3$ grains as the main dust species formed. 
Indeed, a C/O ratio $< 1$ is observed in several low-mass red novae objects (e.g., V1309 Sco; \citealt{Tylenda2016}).

Nevertheless,  we still consider the formation of C grains in our CE ejecta. There have been suggestions that CO molecules can be easily destroyed by energetic electrons, which enables, even for C/O~$< 1$, the condensation of C grains from gas-phase C atoms that are not bound in CO molecules (\citealt{Clayton1999}). In addition, observations of classic novae have revealed the concurrent formation of silicate and carbon grains in their ejecta (\citealt{Evans2005}, \citealt{Sakon2016}). These works indicate that the formation of CO molecules is not always complete, and that C grains can condense in the ejected gas with C/O~$< 1$. Hence, we also address the formation of C grains under the extreme assumption that no CO molecule is formed. Finally, investigating the formation of C grains is applicable to the CE ejecta of asymptotic giant branch (AGB) stars.

For both of the grain species, the amount of dust that forms depends on the abundance of elements that can solidify into dust. Generally for MgSiO$_3$ grains, Si atoms are assumed to combine with O to form SiO molecules, whose abundance is equal to that of the Si atoms. Therefore, the amount of MgSiO$_3$ grains formed is proportional to the least abundant element between Mg and Si. In our case, it is Si, which we will use as the key species for all the calculations related to MgSiO$_3$ grains. For C grains the key species is simply C, which we will use as the key species for the related calculations.

\subsection{Application of the dust model to the common envelope simulation outputs}
\label{ssec:application_of_the_model_to_the_code_data}
The condensation of dust takes place in the cooling gas, where the unsaturated state turns into the supersaturated state determined by the condition $S > 1$, where $S$ is the supersaturation ratio of the gas (equation~56 of \citealt{Nozawa2013}). Thus, to determine if dust  condensation takes place, we need to follow the supersaturation ratio, which depends on the temperature and partial pressure of condensible gaseous atoms \citep{Nozawa2013}.
Our hydrodynamic simulations of the CE ejecta record the density and pressure (which can be converted to temperature), as a function of time for each SPH particle. Hence, the feasibility of dust formation can be assessed, based on the evolution of $S$ calculated from the simulation outputs. 

\citet{Nozawa2013} found that the formation of MgSiO$_3$ and C grains in the expanding gas can be achieved when the non-dimensional quantity $\Lambda_{\rm on} = \tau_{\rm sat}(t_{\rm on})/\tau_{\rm coll}(t_{\rm on})$ is higher than unity, where $\tau_{\rm sat}$ is the time-scale in which the supersaturation ratio increases, $\tau_{\rm coll}$ is the time-scale on which the gas particles collide, and $t_{\rm on}$ is the onset time of dust formation; $\Lambda_{\rm on}$ represents the ratio of time-scales between the seed nuclei formation and grain growth at $t_{\rm on}$. 
Note that the supersaturation ratio increases with decreasing gas temperature: $\ln S \propto T^{-1}$. Hence, $S$ is sensitive to gas temperature $T$, and $\tau_{\rm sat} = |d \ln S/dt|^{-1}$ is approximately proportional to the cooling time of the gas.

\citet{Nozawa2013} also showed that the typical radius of newly formed dust can be determined by the value of $\Lambda_{\rm on}$. In this study, we examine the possibility of dust formation, and estimate the size and mass of dust by calculating $\Lambda_{\rm on}$ for each SPH particle and for the entire duration of the simulation.
In our calculations, the onset time of dust formation $t_{\rm on}$ is defined as the time when $S$ reaches $10$. The gas temperatures when $S = 10$ are $\simeq$1200--1600 K and $\simeq$1700--2100 K for MgSiO$_3$ and C grains, respectively, depending on the number density of the key gas species. Since the time-scale of dust formation (a few days) is much shorter than the dynamical time-scale of the ejecta, $t_{\rm on}$ can be regarded as the condensation time of dust. Dust destruction by sputtering in shocks is not taken into account in the current model, and one might argue that shocks produced during the dynamic in-spiral could potentially destroy part of the dust. However, the erosion of dust by sputtering requires the presence of shocks travelling in the medium at velocities $\gtrsim 100$~km s$^{-1}$ (\citealt{Nozawa2006, Nozawa2007}). Shocks of such magnitude are not present in our CE environment, which exhibits only weak shocks with velocities of a few tens km s$^{-1}$. In addition, we do not see any shock heating after the termination of the dynamic in-spiral, with the ejecta steadily cooling down once the initial interaction is completed (see panel (b) of figure~6 in \citealt{Iaconi2019}). Therefore, shock-driven dust destruction should be negligible.

The formula used by \citet{Iaconi2019} to calculate $\Lambda_{\rm on}$ assumed homologous evolution of the density and temperature of the gas. This assumption affects the value of $\Lambda_{\rm on}$, since the time-scale of gas cooling is regulated by the time-scale of gas expansion. However, our simulation reveals that the evolution of the ejecta can be described by an homologous model only after $\simeq 5000$~days. Here we remove the assumption of homologous expansion for the evolution of density and temperature, and we derive $\Lambda_{\rm on}$ directly and independently for each SPH particle on the basis of the time-scale of gas cooling obtained from the temperature derivative between one code dump and the following one (referred to as non-homologous case).

A comparison between homologous and non-homologous cases reveals as expected a shorter cooling time-scale, on average, in the non-homologous case between $\simeq 2500$~days and $\simeq 5000$~days, due to the faster temperature decrease (panel b of figure~6 in \citealt{Iaconi2019}). At its maximum, the difference in cooling time-scale between the two cases is about a factor of two. This mainly leads to the formation of a larger number of smaller dust grains in the non-homologous case. Nevertheless, we would not expect a large difference in the location of the photosphere because the radius and temperature of the photosphere do not depend greatly on the grain radius so long as it is less than 0.1~$\mathrm{\mu m}$ \citep{Iaconi2019}.

We also stress here that the dust model we use is not limited only to the physical environments of supernovae. It was shown that the model can be applied to dust forming environments expanding slower than supernova ejecta, such as stellar winds (\citealt{Nozawa2014}). In summary, the dust formation model used here has already been validated in a wide range of physical environments (density and temperature of the gas), which also fully cover the physical conditions of the CE ejecta.

\subsection{Impact of the numerical setup on the dust formation model}
\label{ssec:impact_of_the_numerical_setup_on_the_dust_formation_model}
In this section we discuss different numerical factors and assumptions that could affect the results of our dust formation calculation.
\subsubsection{Recombination energy and bound portion of the ejecta}
\label{sssec:recombination_energy_and_bound_portion_of_the_ejecta}
Our simulation does not include radiative heating or cooling, therefore the gas heats and cools adiabatically. However, the hot plasma of the giant's envelope recombines after the ejection and can be an important source of radiative cooling. In this work we model the recombination of the ejecta's gas by using a tabulated equation of state, taken from the MESA code (\citealt{Paxton2011}), which allows the gas of the envelope to recombine when it reaches sufficiently low temperatures. The full payload of the recombination energy released is instantly thermalised and added to the internal energy of the gas, increasing its pressure and allowing it to do work. This results in the unbinding of $94\%$ of the stellar envelope. Whether an SPH particle is unbound is determined by checking if the sum of its kinetic, potential and thermal energies is larger than zero ($E_{\mathrm{kin}} + E_{\mathrm{pot}} + E_{\mathrm{therm}} > 0$).

Here we do not argue on the merits of recombination energy in unbinding the stellar envelope (for that, see references in Section \ref{sec:introduction}). Rather, we use a CE simulation where the gas becomes almost completely unbound as must be the case in at least some cases in Nature. The latent recombination energy in our stellar envelope is in principle sufficient to fully unbind the gian's envelope, if it is fully utilised to do work and does not escape, something that is not the case for similar but even slightly more massive giants (see the calculations in section~7.2 of \citealt{Iaconi2018}). Including recombination energy without including radiation transfer results in the portion of energy that should be radiated away to stay in the gas, increasing its temperature and possibly affecting the timing of dust formation and the size of dust formed (both depending on the gas temperature; see \citealt{Nozawa2013}).  

\citet{Reichardt2020} carry out a detailed analysis of the availability of recombination energy to do work in their numerical setup and conclude that $\simeq 50\%$ of the hydrogen recombination energy and $\simeq 95\%$ of the helium recombination energy are effectively available to do work. Helium recombination is very efficient in driving the envelope because it is released deep in the star. We therefore expect that the unbound mass in our simulation is approximately correct, since only $50\%$ of the energy from recombined hydrogen may be radiated away. 

According to our model, dust formation takes place in neutral, unbound shells. Assuming all unbound ejecta to be neutral, we can estimate the average total energy of a single unbound SPH particle as $E_{\mathrm{kin}} + E_{\mathrm{pot}} + E_{\mathrm{therm}} \simeq 1.65 \times 10^{41}$~erg, where $E_{\mathrm{therm}}$ includes the recombination energy contributions to the internal energy from H and He (again, see \citealt{Reichardt2020} for details). Note that such contributions are the recombination energies values before having subtracted the estimated amounts that have been radiated away. By using the numbers provided by \citet{Reichardt2020} at $359$~days from the beginning of the simulation ($\simeq 50$~days after the end of the dynamic in-spiral), we estimate the energy lost via radiative cooling of a single unbound SPH particle to be $\simeq 4.2 \times 10^{40}$~erg. If we compare the two numbers we can see that about $25\%$ of the total energy would be lost via radiative cooling. Without including it, the ejecta will therefore reach dust formation temperatures later, at lower densities, and our dust model might be producing dust grains slightly smaller than those obtained in a model including radiation transfer. 

The bound portion of the envelope is very small and, except for a few SPH particles that remain close to the central binary, the remaining ones are mixed up with the unbound ejecta. These particles are very loosely bound and behave similarly to their unbound neighbours, therefore exhibiting a similar pattern of dust formation.

\subsubsection{The temperature of dust and gas}
\label{sssec:the_temperature_of_dust_and_gas}
Being unable to calculate the dust temperature, we have assumed that it is the same as the gas': $T_{\mathrm{dust}} = T_{\mathrm{gas}}$. Here we consider the validity of this assumption for the formation of the seed grains and for grain growth.

We started by estimating the collisional heating rate, $R_{\mathrm{coll}}$, the radiative heating rate, $R_{\mathrm{heat}}$, and the radiative cooling rate, $R_{\mathrm{cool}}$, by applying the prescription of \citet[their equation~1]{Nozawa2008} to our CE ejecta during the dust formation period. The three rates determine $T_{\mathrm{dust}}$ via the balance equation:

\begin{equation}
R_{\mathrm{coll}} + R_{\mathrm{heat}} = R_{\mathrm{cool}} \ . 
\end{equation}
\label{eq:dust_rates}

The values we obtain are $R_{\mathrm{coll}} \simeq 7 \times (10^2$ --- $10^4) \chi$~erg cm$^{-2}$ s$^{-1}$, $R_{\mathrm{heat}} \simeq 3.5 \times (10^5$ --- $10^6) \chi$~erg cm$^{-2}$ s$^{-1}$ and $R_{\mathrm{cool}} \simeq 1.1 \times (10^7$ --- $10^8) \chi$~erg cm$^{-2}$ s$^{-1}$, where $\chi$ is the geometrical cross section of the dust particles. With the typical physical quantities of the post-CE ejecta considered here, the energy involved in collisional heating between dust and gas is negligible compared to that involved in radiative heating/cooling of dust. The temperature of dust is higher or lower than that of the gas, respectively, if radiative heating dominates or if radiative cooling dominates. In particular, small dust grains are susceptible to rapid temperature fluctuations.

In terms of the formation of seed grains, we can safely assume that $T_{\mathrm{dust}} = T_{\mathrm{gas}}$. In fact, the seed grains are very small ($< 30$ atoms) and can be regarded as large molecules, therefore sharing a similar temperature to that of the gas where they form. That said, \citet{Keith2011} claim that the formation rate of seed grains can be enhanced by temperature fluctuations. However, they consider physical conditions very different from those present in astronomical environments (very high temperatures and very high gas densities), therefore the results of their study cannot simply be included in the dust formation model we used here. If temperature fluctuations were present and resulted in variations of the formation rate of seed grains, we would expect the onset of dust formation to be shifted at earlier/later times.
Moreover, \citet{Paquette2011}, systematically analysed the effects of the formation rate of seed grains on their final size, and found that a substantial increase in the former results only in a small increase in the latter. Therefore, even if temperature fluctuations would affect the formation rate of seed grains, it would not have significant effects on our results for the grain size. 

Our assumption that dust and gas temperatures are the same, could be more of an issue for grain growth. Dust temperatures higher than gas temperatures due to radiative heating might lead to evaporation of the grains and effectively quench grain growth. We can observe that in general $R_{\mathrm{heat}} < R_{\mathrm{cool}}$ by one order of magnitude, and a more accurate calculation showed that this is true for all the SPH particles for both C and MgSiO$_3$ grains. This is more easily satisfied by MgSiO$_3$ because the opacity (absorption efficiency) of MgSiO$_3$ is lower at the black-body peak wavelength of the central object. Therefore, radiative heating falls below radiative cooling, meaning that dust temperature must be always lower than gas temperature. This may lead to an increase in grain growth rate, but in our model we assume the sticking probability to be constant. Therefore, in our formalism, as the temperature of dust decreases the grain growth rate does not increase. It is difficult to model grain growth by accounting for the effect of temperature fluctuations and to predict what the behaviour would be under the present circumstances if such effects were taken into account. We leave such an investigation to future work.

We additionally note that since $R_{\mathrm{heat}} < R_{\mathrm{cool}}$, the radiation from the central object has a negligible effect on dust destruction and that the radiation pressure generated would accelerate the dust to only a few tens km s$^{-1}$, therefore the effect on the ejecta dynamics would be small.


\section{Dust formation process}
\label{sec:dust_formation_process}
\begin{figure*}
\centering     
\subfigure[]{\includegraphics[scale=0.45, trim=0.0cm 0.0cm 0.0cm 0.0cm]{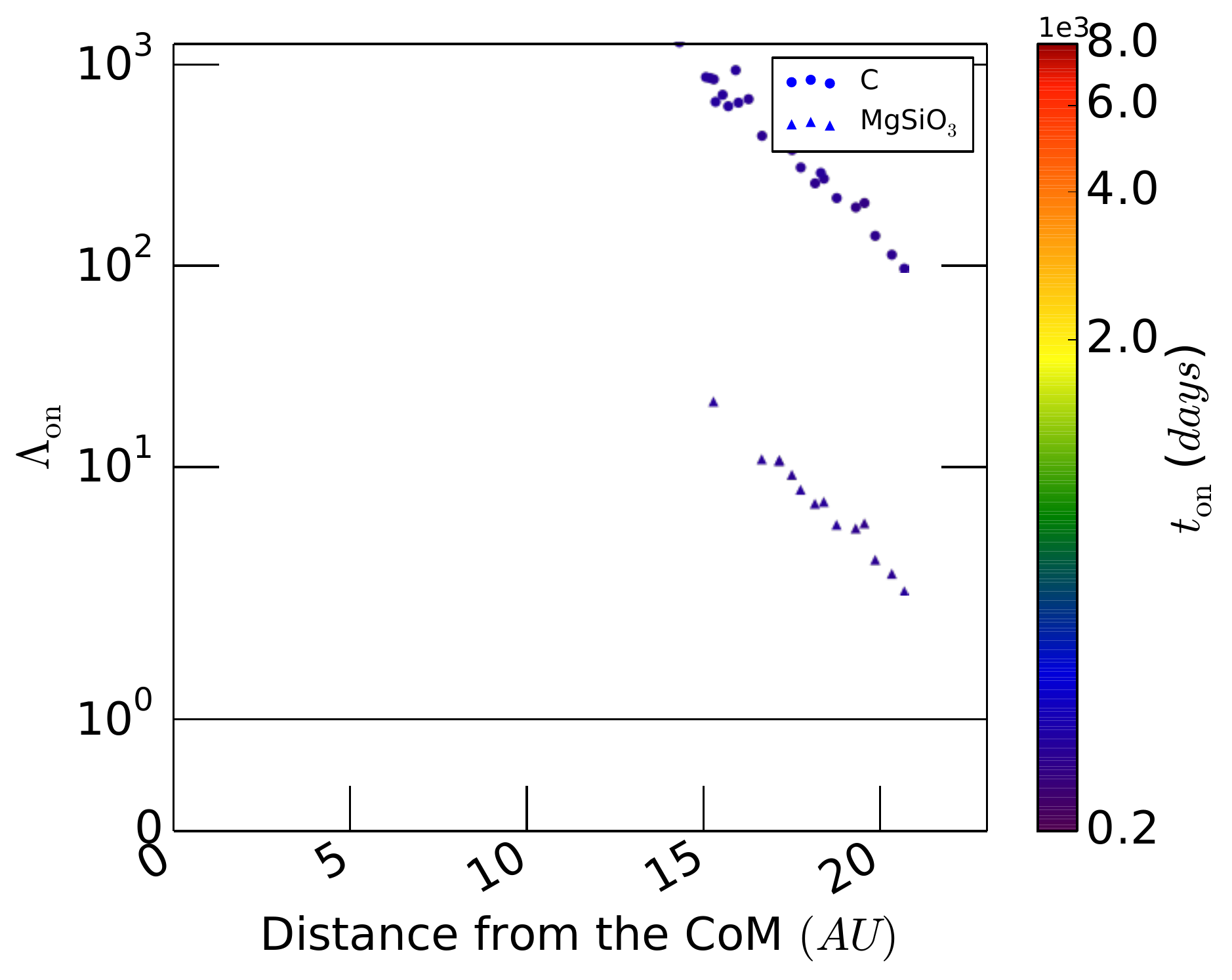}}
\subfigure[]{\includegraphics[scale=0.45, trim=0.0cm 0.0cm 0.0cm 0.0cm]{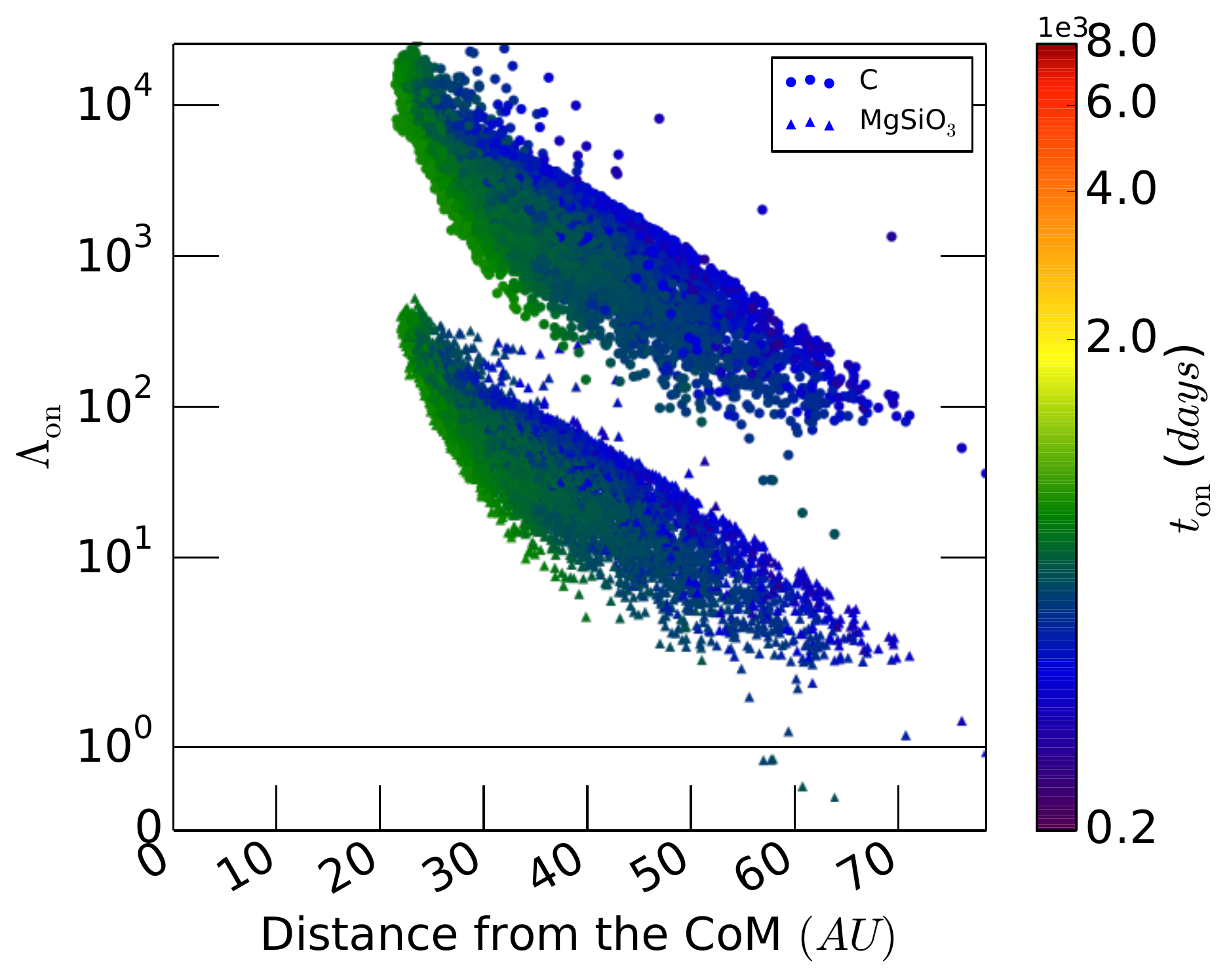}}
\subfigure[]{\includegraphics[scale=0.45, trim=0.0cm 0.0cm 0.0cm 0.0cm]{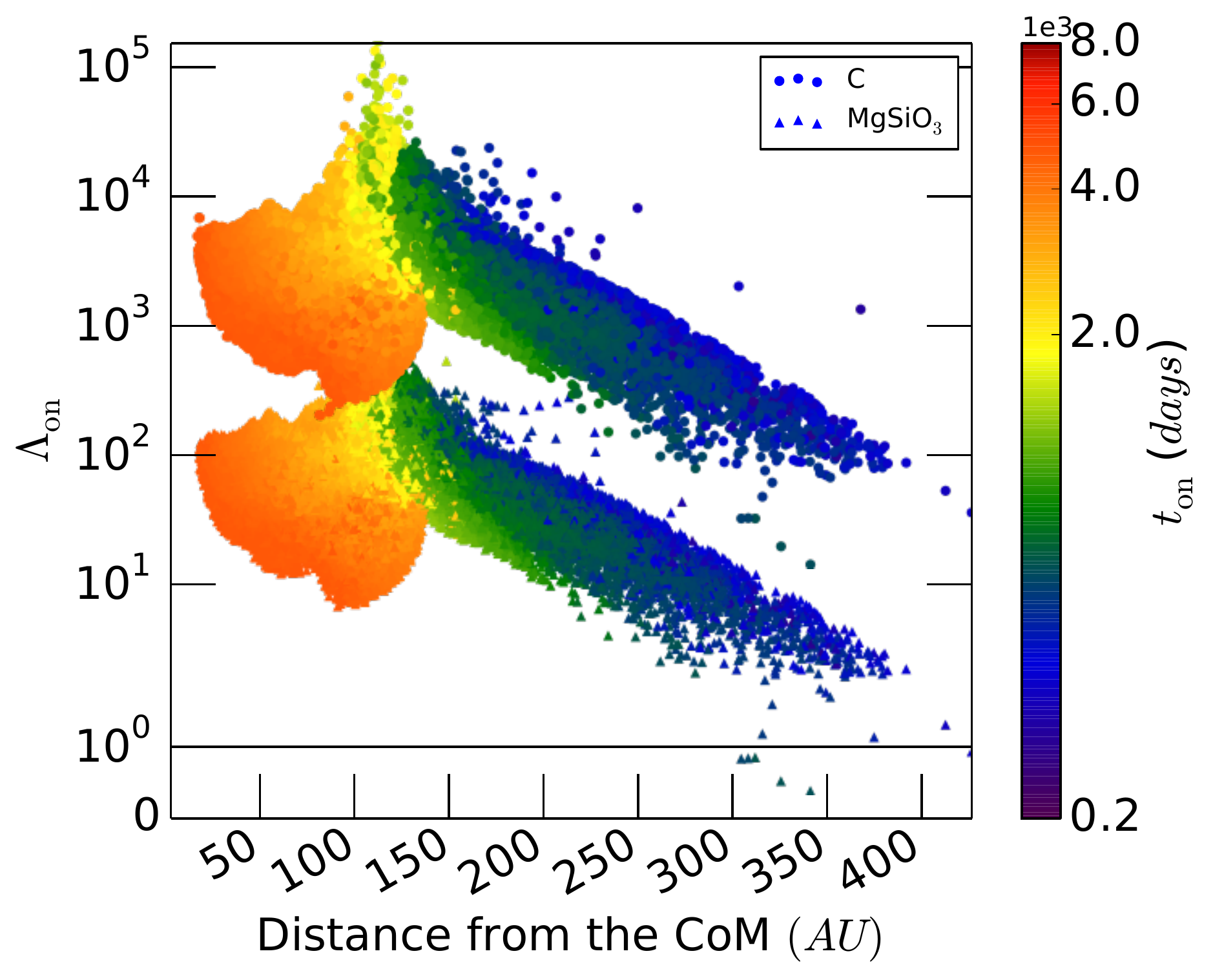}}
\subfigure[]{\includegraphics[scale=0.45, trim=0.0cm 0.0cm 0.0cm 0.0cm]{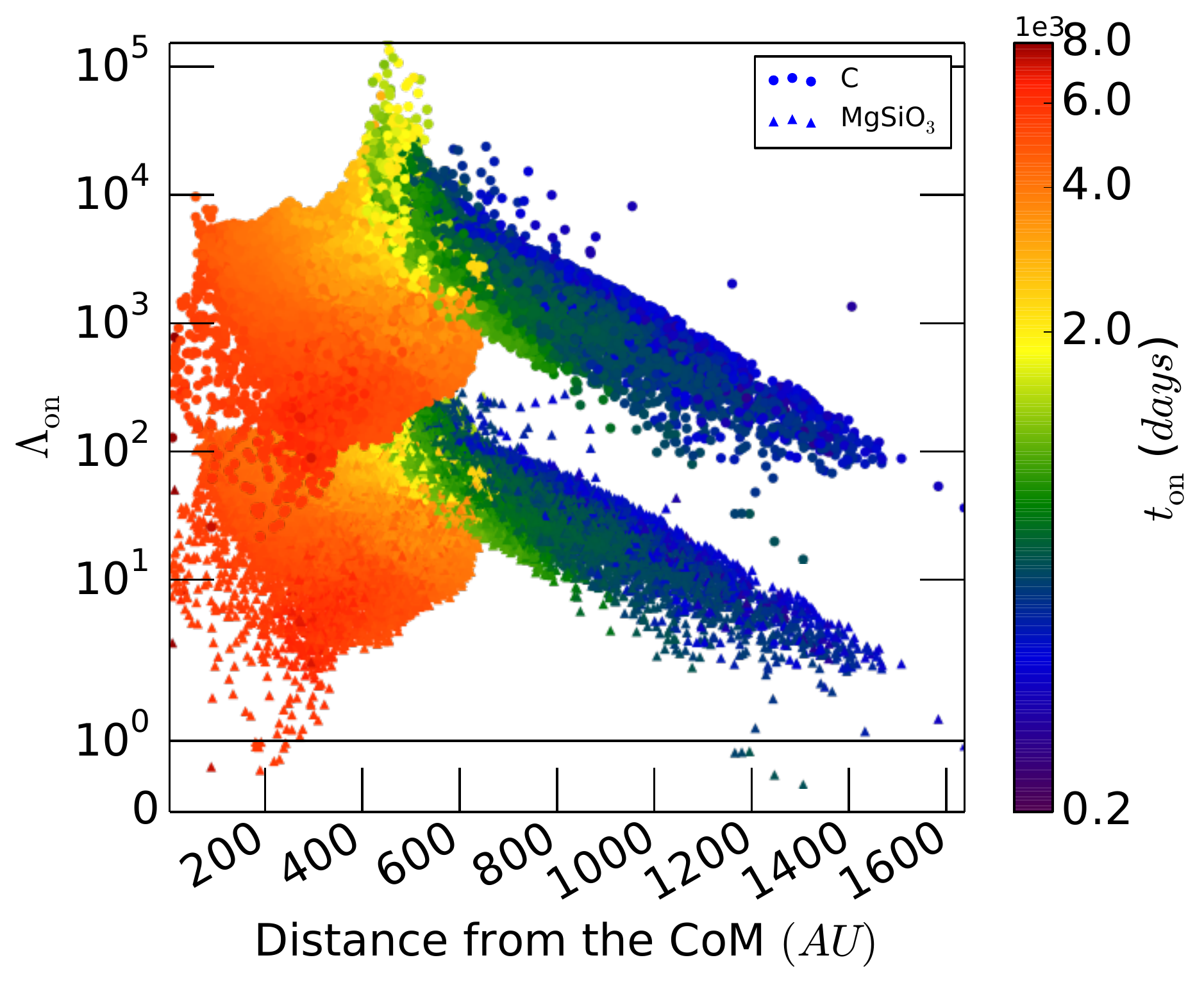}}
\caption{\protect\footnotesize{$\Lambda_{\rm on}$ for the particles that have achieved the condition $S \geq 10$ as a function of the distance from the CoM of the system, at $\simeq 300$~days (a), $\simeq900$~days (b), $\simeq5000$~days (c) and $18434$~days (d) for C and MgSiO$_3$ grains. The colour scale represents $t_{\rm on}$, while the black horizontal line represents $\Lambda_{\rm on} = 1$.}}
\label{fig:lambda_on}
\end{figure*}

In Figure~\ref{fig:lambda_on} we plot $\Lambda_{\rm on}$ as a function of the distance from the center of mass (CoM) of the system and for both types of dust considered in this work. The particles shown in each panel are only those that have achieved the condition for the onset of dust formation ($S \geq 10$) before the time at which each snapshot is taken.
We plot these quantities at four different times: the end of the dynamic in-spiral ($\simeq 300$~days), the end of the formation of the outer dust population ($\simeq 900$~days; see below and Section~\ref{ssec:locations_of_the_grains}), the onset of homologous expansion in the ejecta ($\simeq 5000$~days) and the end of the simulation (18\,434~days). Additionally, we show as a colour gradient the onset time of dust formation, $t_{\mathrm{on}}$, of the SPH particles.

For the great majority of the particles, the values of $\Lambda_{\rm on}$ we record at $t_{\rm on}$ are larger than unity (in Figure~\ref{fig:lambda_on} all the SPH particles for C grains and most of them for MgSiO$_3$ grains, respectively, reside above the black horizontal line). This means that, for the great majority of the particles, the nucleated seed clusters can grow stably through the accretion of the key gas species. Therefore, nearly the maximum possible amount of dust is formed in our simulation. The difference in the values of $\Lambda_{\rm on}$ between the two dust types mainly stems from the difference in condensation temperature and number abundance of the key gas species.

We observe two distinct populations for $\Lambda_{\rm on}$: an outer population represented by the diagonal strip formed between $\simeq 300$ and $\simeq 900$~days in the outer layer and an inner population formed by the clump that forms in the inner layers starting from $\simeq 900$~days. We will analyse the two populations in detail in Sections~\ref{ssec:locations_of_the_grains} and \ref{ssec:grain_size_distribution}, when discussing the locations where dust forms and the grain size distribution.

In the outer dust population, the formation of grains mostly proceeds from the outside-in. We notice however that for a given distance, dust forms at different times. For example, by picking a distance of $800$~AU in Figure~\ref{fig:lambda_on} (panel (d)), we can see that the colour changes from blue in the upper part of the distribution of each dust type to green in the lower part, which corresponds to a spread in time of $\simeq 900$~days. In the inner dust population, we observe instead a much larger spread of $t_{\rm on}$ at a fixed distance. For example, if we consider a distance of 400~AU in Figure~\ref{fig:lambda_on} (panel (d)), the range of $t_{\rm on}$ is of the order of several thousands of days. Therefore, especially for the inner dust population, regions that are equidistant from the CoM see the coexistence of dusty and dust-free gas. This is due to the complex dynamic and thermodynamic interactions taking place in the inner portions of the ejecta even after the dynamic in-spiral (\citealt{Iaconi2019}). 

\begin{figure*}
\centering     
\subfigure[]{\includegraphics[scale=0.45, trim=0.0cm 0.0cm 0.0cm 0.0cm, clip]{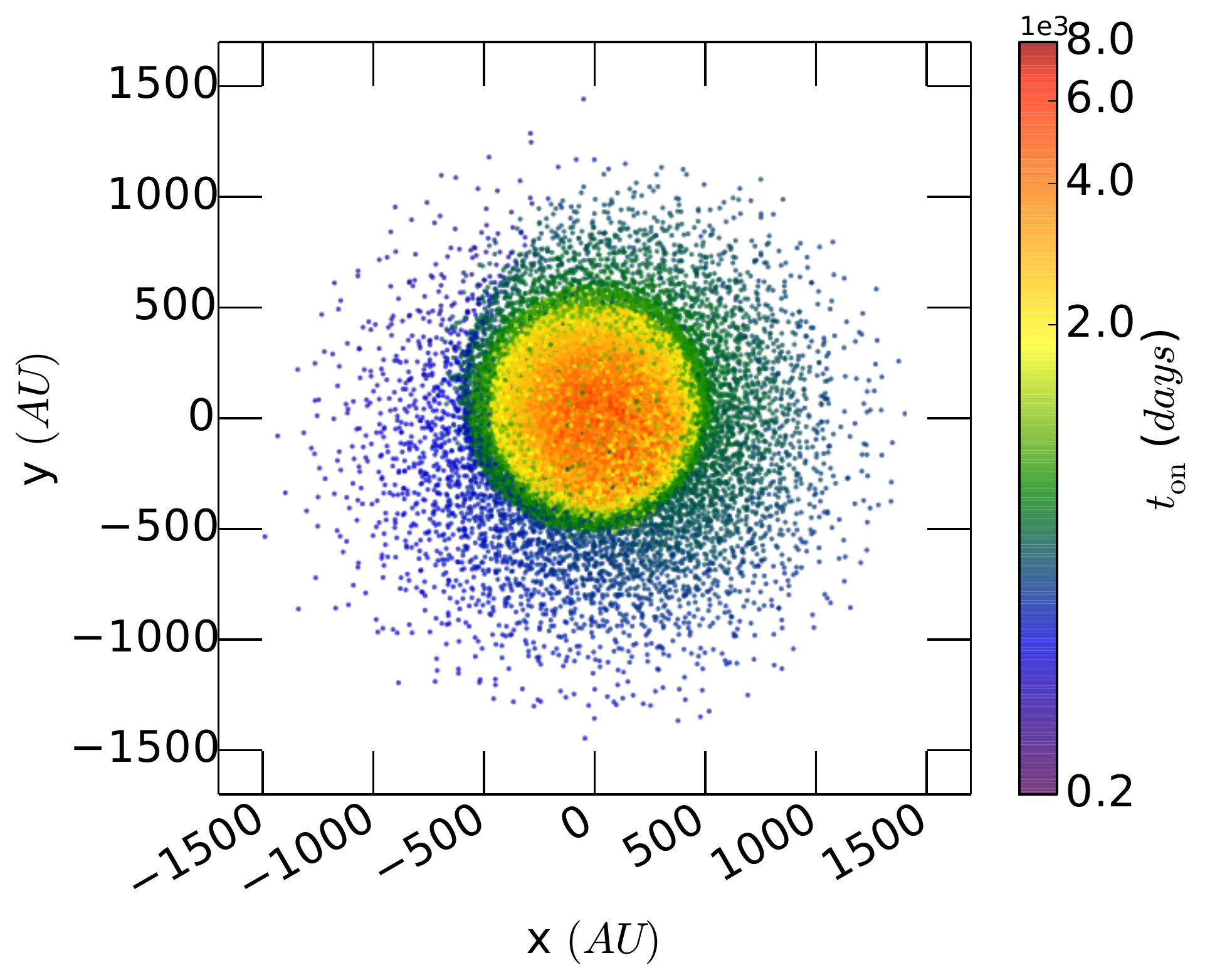}}
\subfigure[]{\includegraphics[scale=0.45, trim=0.0cm 0.0cm 0.0cm 0.0cm, clip]{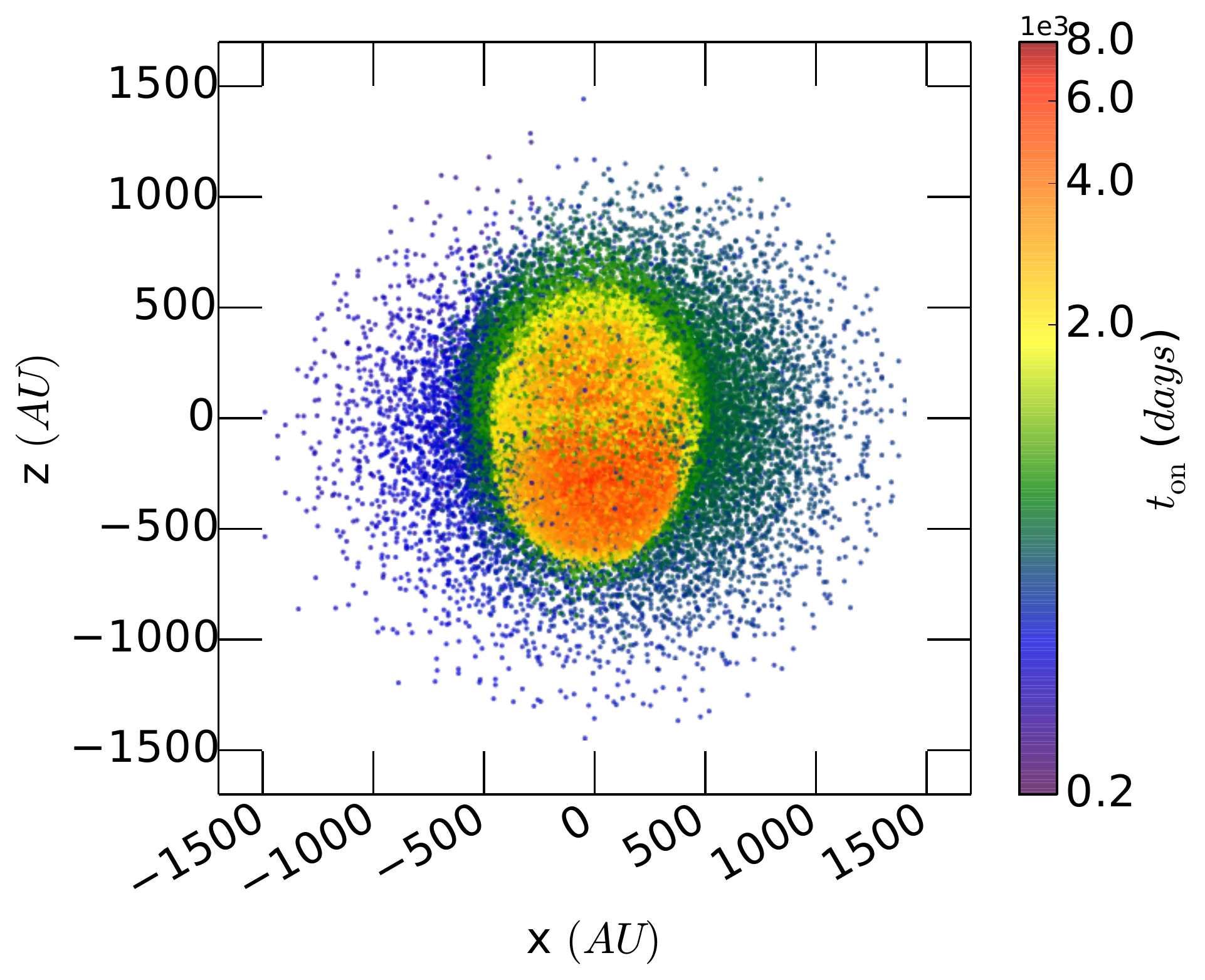}}
\caption{\protect\footnotesize{Projections on the $x-y$ (panel a) and $x-z$ (panel b) planes of the SPH particles that formed C grains, $t_{\mathrm{on}}$ is represented by the colour scale. The projections are taken at $18434$~days, the end of the simulation. The results for MgSiO$_3$ grains are very similar to those showed here.}}
\label{fig:onset_time_map_19999}
\end{figure*}

To clarify how $t_{\mathrm{on}}$ depends on the location, we plot in Figure~\ref{fig:onset_time_map_19999} the projection of SPH particles that formed C grains at $18\,434$~days on the x-y (panel (a)) and x-z (panel (b)) planes, where the colour scale represents $t_{\mathrm{on}}$. 
The results for MgSiO$_3$ are very similar to those for C and we do not show them.
In the projections it is clear that the outer portion of the ejecta has a more evident $t_{\mathrm{on}}$ gradient, while the inner region has a more mixed $t_{\mathrm{on}}$ distribution.

The process of dust formation, which starts right after the completion of the dynamic in-spiral, lasts until $\simeq 5000$~days. This corresponds to the time required for the ejecta to achieve homologous expansion (\citealt{Iaconi2019}). At $\simeq 5000$~days $\simeq 96\%$ and $\simeq 97\%$ of the particles, for MgSiO$_3$ and C, respectively, have formed dust. Most of the remaining particles slowly form dust over the remaining $\simeq 15\,000$~days of the simulation. The fact that the dust formation process starts at the end of the dynamic in-spiral and terminates when the ejecta achieve homologous expansion is purely coincidental.



\section{Characteristics of the dust}
\label{sec:characteristics_of_the_dust}
The model proposed by \citet{Nozawa2013} allows us to evaluate several properties of the newly formed dust. In this section we will analyse these properties and their implications.

\subsection{Average grain size}
\label{ssec:average_grain_size}
\begin{figure*}
\centering     
\includegraphics[scale=0.45, trim=0.0cm 0.0cm 0.0cm 0.0cm, clip]{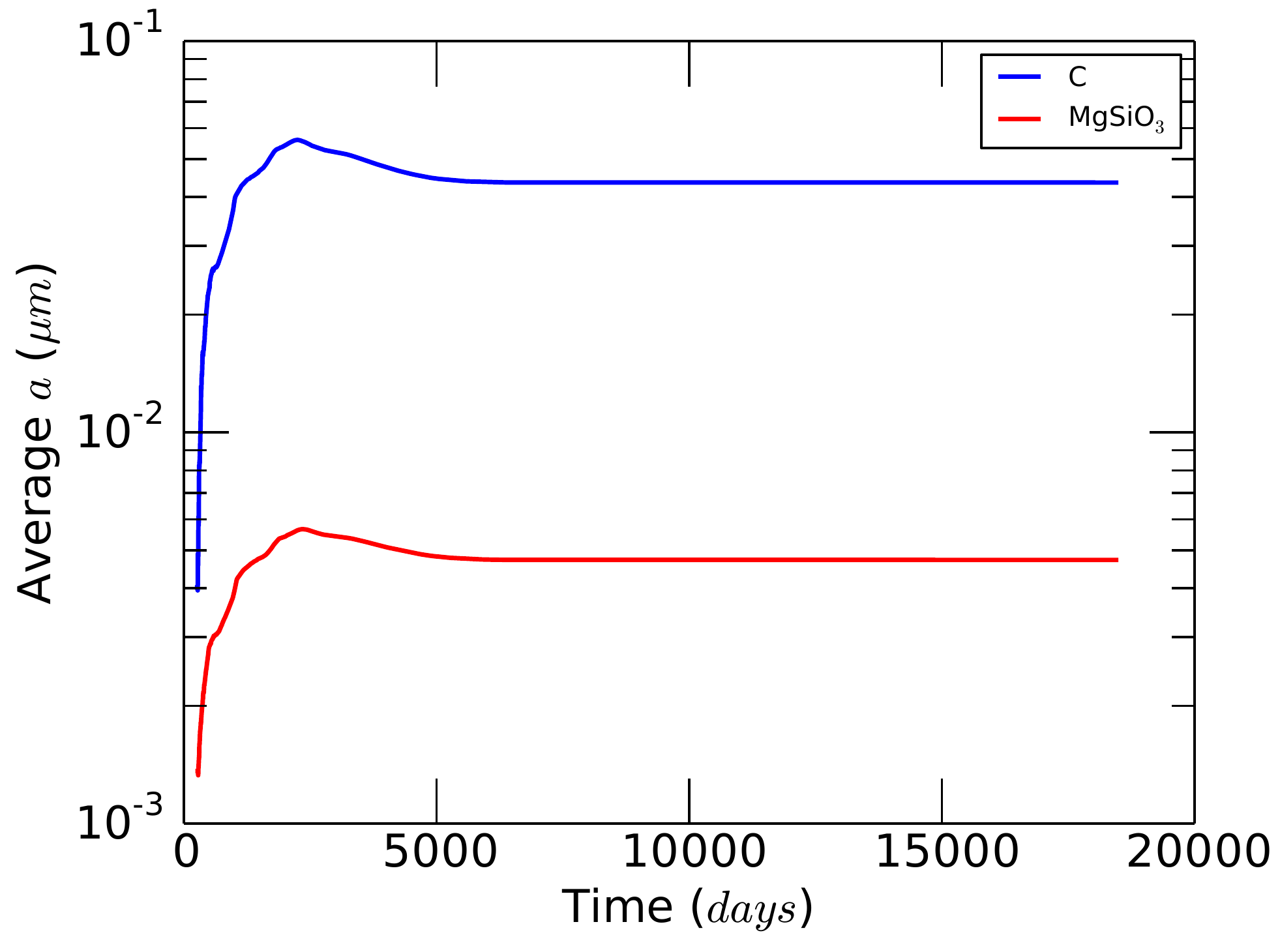}
\caption{\protect\footnotesize{Average grain size, $a_{\mathrm{ave}}$, for MgSiO$_3$ and C dust as a function of time.}}
\label{fig:alpha_ave_vs_time}
\end{figure*}

Let us first consider the evolution of the average grain radius, $a_{\mathrm{ave}}$, as a function of time (Figure~\ref{fig:alpha_ave_vs_time}).
For both MgSiO$_3$ and C grains we observe a general increase in $a_{\mathrm{ave}}$ during the first $\simeq 2500$~days of the simulation. This is followed by a small decrease up to $\simeq 5000$~days, when all the possible dust grains have formed (Section~\ref{sec:dust_formation_process}) and $a_{\mathrm{ave}}$ becomes constant.

Larger grains form on average at later times. In the model we use, $a_{\mathrm{ave}}$ is roughly proportional to $\Lambda_{\rm on} \propto \rho_{\mathrm{on}} T_{\mathrm{on}}^{3/2}$ (but see equation~64 of \citealt{Nozawa2013} for the correct relationship), where $\rho_{\mathrm{on}}$ is the density of an SPH particle at the onset of dust formation and $T_{\mathrm{on}}$ is its temperature (equations~63 and 64 of \citealt{Nozawa2013}). Hence an increase in the average grain size means that dust grains form in denser layers of the ejecta at later times.
In Figure~\ref{fig:lambda_on} (panel (d)) we observe that, within 600~AU from the CoM of the system, the maximum value of $\Lambda_{\rm on}$ is approximately constant. This region becomes populated between $900$ and $5000$~days. A constant maximum $\Lambda_{\rm on}$ results in a constant maximum grain size. As a result the values of $a_{\mathrm{ave}}$ change only depending on the amount and size of smaller dust grains. The increase in number of smaller grains between $900$ and $5000$~days reduces the value of $a_{\mathrm{ave}}$ and produces the small bump we observe after the initial steady increase.

The difference in average radius between C and MgSiO$_3$ grains is instead dictated by the different number densities of the key elements of the two dust types, C and MgSiO$_3$, respectively. The former has a higher number density, resulting in a final $a_{\mathrm{ave}} \simeq 4 \times 10^{-2}$~$\rm \mu m$, while the latter has a lower one, and a final $a_{\mathrm{ave}} \simeq 5 \times 10^{-3}$~$\rm \mu m$.

\subsection{Location of the grains}
\label{ssec:locations_of_the_grains}
\begin{figure*}
\centering     
\subfigure[]{\includegraphics[scale=0.4, trim=0.0cm 0.0cm 0.0cm 0.0cm, clip]{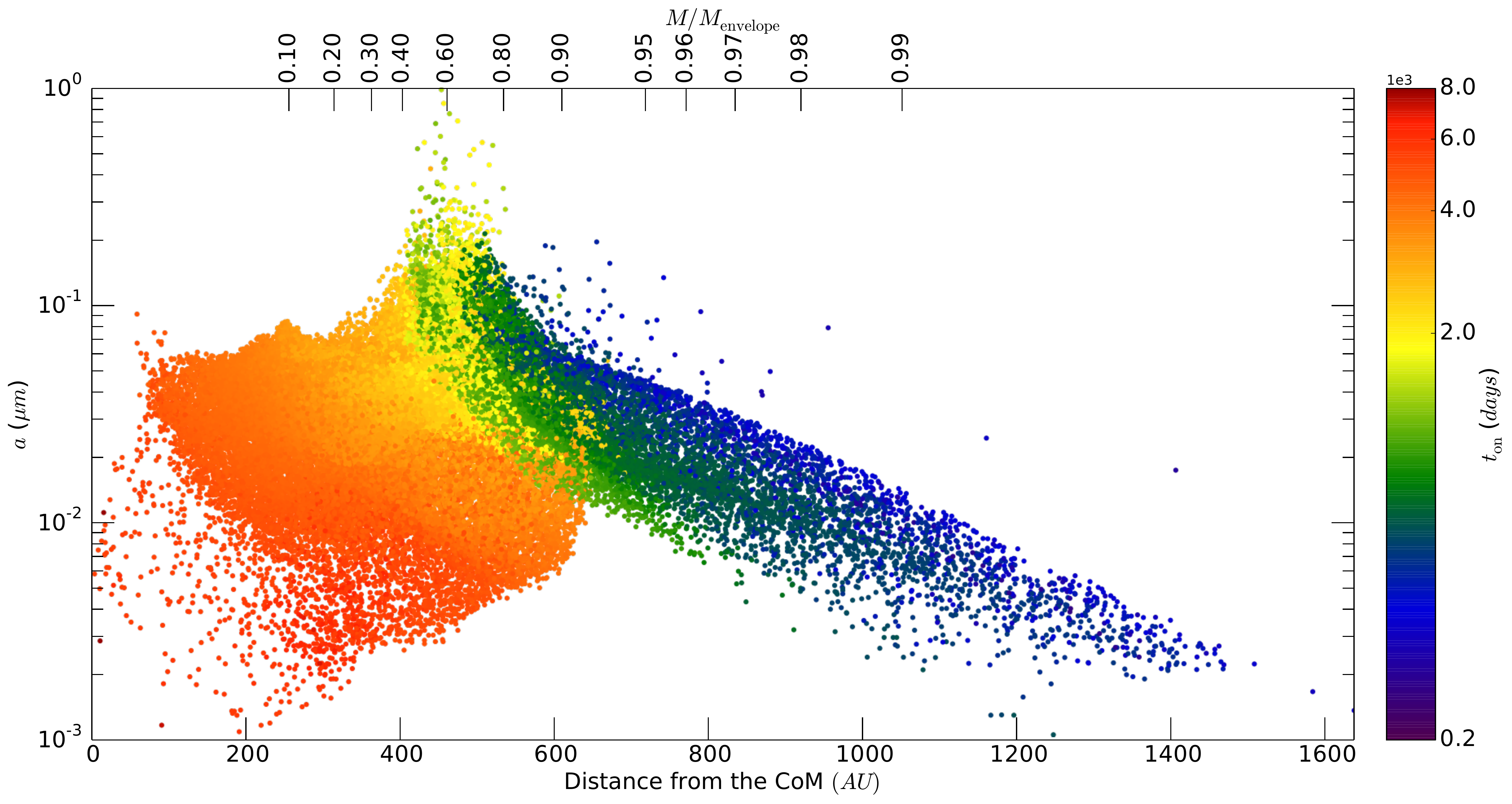}}
\subfigure[]{\includegraphics[scale=0.4, trim=0.0cm 0.0cm 0.0cm 0.0cm, clip]{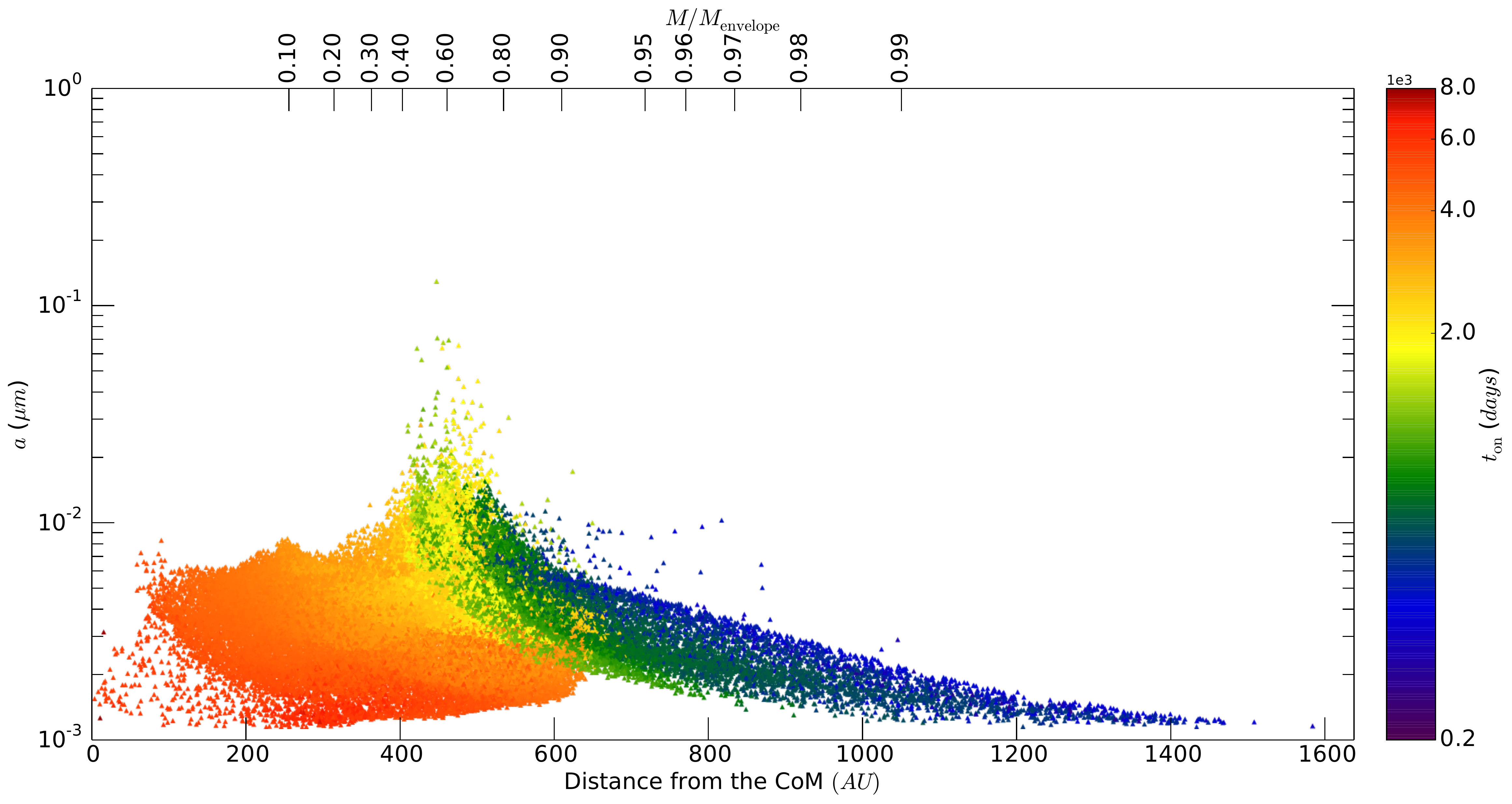}}
\caption{\protect\footnotesize{Grain radius, $a$, as a function of the distance from the CoM of the system for C (a) and MgSiO$_3$ (b) grains at the end of the simulation. On the upper x axis we plot the mass coordinate with respect to the total ejecta mass. The colour scale represents $t_{\rm on}$.}}
\label{fig:alpha_vs_radius}
\end{figure*}

By looking at the distribution of grains with respect to the CoM of the system (Figure~\ref{fig:alpha_vs_radius}) we can see two distinct populations. The outer dust population is composed by the grains that form earlier and shows a decreasing grain size at increasing distance from the CoM. In Figure~\ref{fig:alpha_vs_radius} the outer population can be identified with the blue to green points that occupy the area between $\simeq 600$~AU and $\simeq 1600$~AU from the CoM. The inner dust population is instead formed by the grains residing at distances $<600$~AU. The two populations form from $\simeq 300$ to $\simeq 900$~days (outer population) and $\simeq 900$ to $\simeq 5000$~days (inner population), and can be correlated with the shapes of the profiles of temperature and density of the ejecta. At the end of the simulation we in fact observe that the region between $\simeq 600$~AU and $\simeq 1600$~AU corresponds to a region where temperature and density decrease at increasing distance from the CoM (for the density plot, see figure~3 of \citealt{Iaconi2019}).

The formation of two populations is due to the envelope ejection dynamics. The SPH particles that produce the outer dust population amount to $\simeq 10\%$ of the ejecta mass and coincide to the part of the ejecta that is unbound purely by being accelerated above escape velocity by the deposited orbital energy (\citealt{Iaconi2017}, \citealt{Iaconi2018}). This part of the ejecta cools down almost adiabatically after being unbound, because it is free to expand into the empty space. As a result the gas temperature decreases
until it reaches values suitable for dust formation at earlier times than in the remaining part of the ejecta.

The inner layers of the ejecta interact with each other in the course of the first $\simeq 5000$~days of the simulation (\citealt{Iaconi2019}). This results in a more complex evolution and produces the inner dust population whose formation time and size are diverse. For both populations, but more so for the second one, grains of different sizes form at different times for a given distance from the centre. This is reminiscent of the behaviour of $\Lambda_{\rm on}$. In fact, in accordance to the formulation of \citet{Nozawa2013}, $a_{\mathrm{ave}} \propto \Lambda_{\rm on} \propto \tau_{\rm sat}(t_{\rm on})/\tau_{\rm coll}(t_{\rm on})$. Therefore, the size of the grains that form in a specific SPH particle depends on the balance between how fast the new seed nuclei form ($\tau_{\rm sat}$) and how fast the gas particles can collide with the seed nuclei and pre-formed grains to grow them ($\tau_{\rm coll}$). The combination of large $\tau_{\rm sat}$ and small $\tau_{\rm coll}$ results in a rapid build up of large grains. As a result, we observe that, at the same distance from the CoM, SPH particles form larger grains at earlier times as well as smaller grains at later times (see also Section~\ref{ssec:average_grain_size}). This process takes place roughly in two phases. First, a front of dust formation, starting from the outskirts of the ejecta, moves inwards and forms the larger grains. 
At this point, in the inner layers of the ejecta we have SPH particles where larger dust grains have already formed and SPH particles where the gas has not yet met the physical conditions for dust formation. The latter are where the smaller dust grains are gradually formed. This process is clear by looking at the distribution of the colours in Figure~\ref{fig:alpha_vs_radius}.



When dust starts forming after the dynamic in-spiral, at $\simeq 300$~days, the first grains form at $\simeq 20$~AU from the CoM of the binary. By $\simeq 5000$~days the dust stops forming. The last grains form at $\simeq 100$~AU. Therefore, we observe dust forming roughly in a 100~AU shell around the binary system. Previous results on dust formation in CE (\citealt{Lu2013}) using a 1D model, estimated a range between $\simeq 7$ and $7 \times 10^{4}$~AU, much larger than the one we obtain here, implying that their ejecta rarefy and cool down more slowly in the 1D model than in our 3D model. 

\subsection{Grain size distribution}
\label{ssec:grain_size_distribution}
\begin{figure*}
\centering     
\subfigure[]{\includegraphics[scale=0.4, trim=0.0cm 0.0cm 0.0cm 0.0cm, clip]{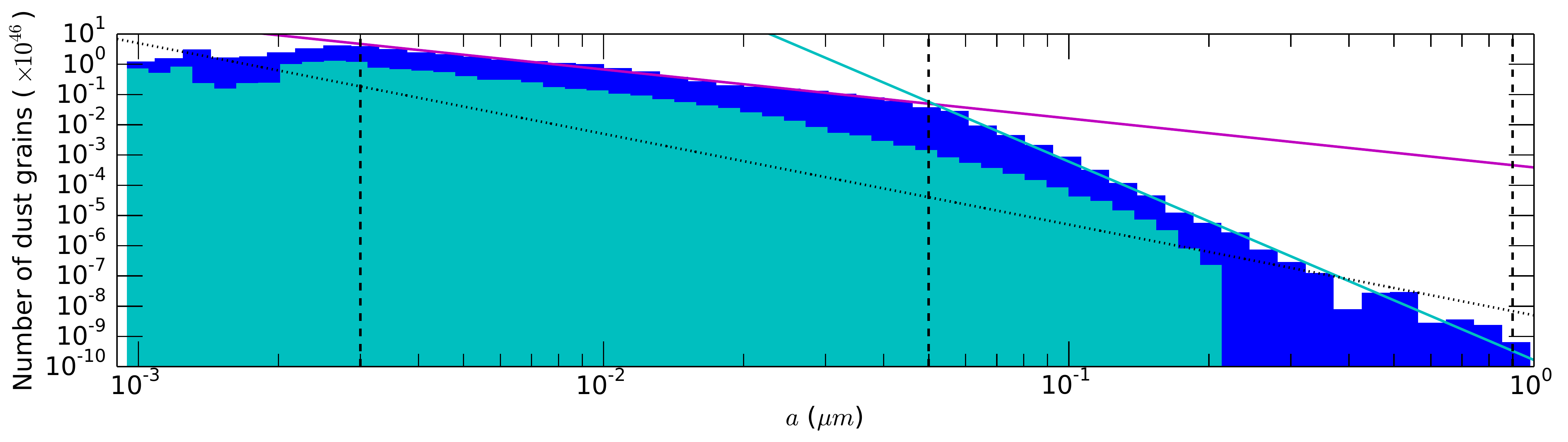}}
\subfigure[]{\includegraphics[scale=0.4, trim=0.0cm 0.0cm 0.0cm 0.0cm, clip]{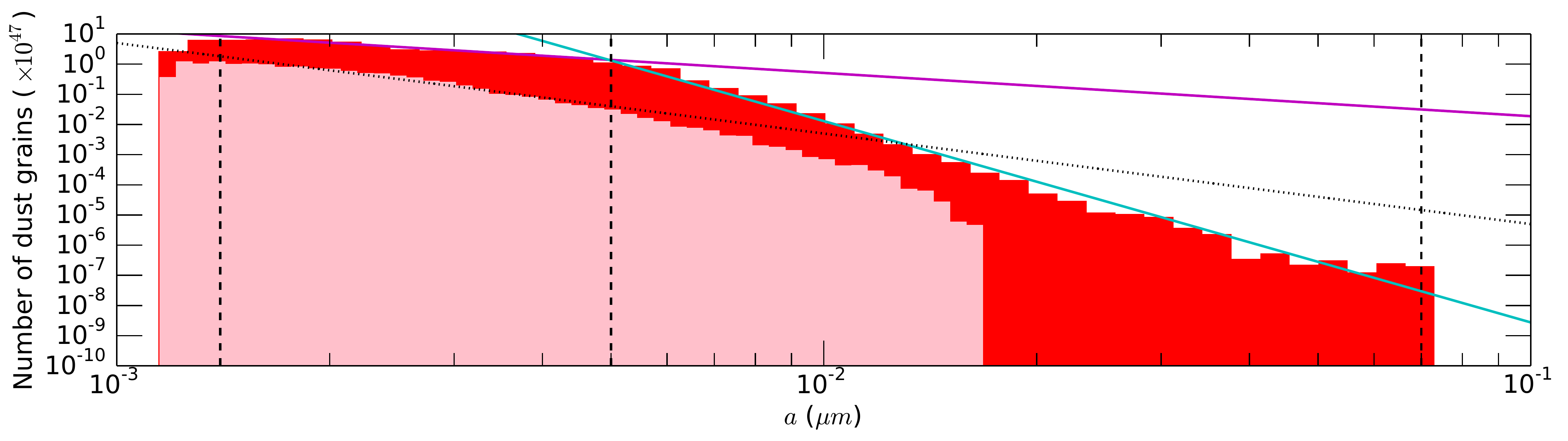}}
\caption{\protect\footnotesize{Number of dust grains as a function of the grain radius, $a$, at the end of the simulation for C grains (a) and MgSiO$_3$ grains (b). In panel (a) the magenta line represents the best fit to a power law in the range $3 \times 10^{-3} \leq a < 5 \times 10^{-2}$~$\mathrm{\mu m}$, the cyan line in the range $5 \times 10^{-2} \leq a \leq 9 \times 10^{-1}$~$\mathrm{\mu m}$. In panel (b) the magenta line represents the best fit to a power law in the range $1.4 \times 10^{-3} \leq a < 5 \times 10^{-3}$~$\mathrm{\mu m}$, the cyan line in the range $5 \times 10^{-3} \leq a \leq 7 \times 10^{-2}$~$\mathrm{\mu m}$. The boundaries of the fitting are marked by the vertical dashed lines, while the dotted line represents a line at constant mass for reference.
We overplot the distribution for the outer dust population (i.e., at $900$~days) in cyan (a) and pink (b), for C and MgSiO$_3$, respectively. Note that the bin size use for the outer dust population is smaller than that used for the data at the end of the simulation.}}
\label{fig:npart_vs_alpha}
\end{figure*}

For both the dust types considered, the size distributions are weighted towards small grains. They peak at $\simeq 3 \times 10^{-3}$~$\mathrm{\mu m}$ for C and $\simeq 1.8 \times 10^{-3}$~$\mathrm{\mu m}$ for MgSiO$_3$ grains. Moreover, we can see that the size of the grains spans a relatively broad range, between $\simeq 3 \times 10^{-3}$ and $\simeq 9 \times 10^{-1}$~$\mathrm{\mu m}$ for C and between $\simeq 1.4 \times 10^{-3}$ and $\simeq 7 \times 10^{-2}$~$\mathrm{\mu m}$ for MgSiO$_3$. This is shown in Figure~\ref{fig:npart_vs_alpha}, where we plot the number of dust grains vs. the grain radius, $a$, for C (panel (a)) and MgSiO$_3$ (panel (b)). 

\begin{figure*}
\centering     
\includegraphics[scale=0.4, trim=0.0cm 0.0cm 0.0cm 0.0cm, clip]{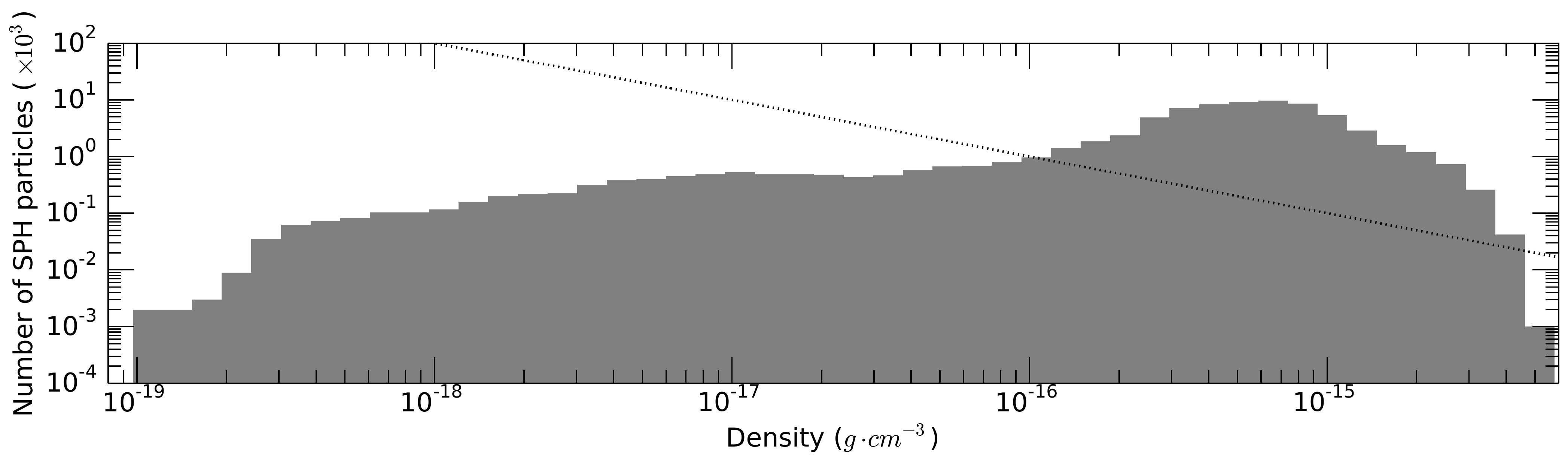}
\caption{\protect\footnotesize{Number of SPH particles as a function of density at the end of the simulation. The dotted line represents a line at constant mass for reference.}}
\label{fig:npart_vs_density}
\end{figure*}

We observe that the distributions have a distinct shape, with a net change of slope at $\simeq 5 \times 10^{-2}$~$\mathrm{\mu m}$ and $\simeq 5 \times 10^{-3}$~$\mathrm{\mu m}$ for C and MgSiO$_3$ grains, respectively. This turnover point, in terms of how much ejecta mass is contained in a single $\Delta a$ bin, represents the average grain size. In other words, the $\Delta a$ bins around the turnover point are those that contain the highest amount of gas mass per number of dust grains. We can match the turnover point in grain size with the maximum in the density distribution of SPH particles (Figure~\ref{fig:npart_vs_density}), located at $\simeq 7 \times 10^{-16}$~g$\cdot$cm$^{-3}$. By looking at the dotted lines in the two panels of Figure~\ref{fig:npart_vs_alpha}, that represent constant masses, it is possible to see that on the left of the turnover points, the magenta lines move closer to the constant mass lines, while on the right of the turnover points the cyan lines moves closer to the constant mass lines. A similar behaviour can be observed for the maximum of the density distribution in Figure~\ref{fig:npart_vs_density}. This informs us of the fact that a turnover point in density results in a turnover point in the grain size distribution.

We are able to fit two power laws through the data, marked by the magenta and cyan lines in Figure~\ref{fig:npart_vs_alpha}. The two power laws have been fit through the data between the vertical dashed lines and the fitting parameters are shown in Table~\ref{tab:fit_parameters}, where for the power law we use the nomenclature $n(a) = Ca^{-q}$, the same as that of \citet{Nozawa2016}. Let us now consider separately the three areas of Figure~\ref{fig:npart_vs_alpha} separated by the vertical dashed lines.

\begin{table*}
\begin{center}
\begin{tabular}{c|ccc|ccc}
Dust type & \multicolumn{3}{c|}{Magenta} & \multicolumn{3}{c}{Cyan} \\
& parameters & & std deviation & parameters & & std deviation \\
\hline
C & \makecell{$C=3.88\times10^{-4}$ \\ $q=1.62$} & \makecell{$\pm$ \\ $\pm$} & \makecell{$6.71\times10^{-5}$ \\ $3.87\times10^{-2}$} &
\makecell{$C=1.65\times10^{-10}$ \\ $q=6.56$} & \makecell{$\pm$ \\ $\pm$} & \makecell{$5.55\times10^{-11}$ \\ $1.92\times10^{-1}$} \\
\hline
MgSiO$_3$ & \makecell{$C=6.92 \times 10^{-4}$ \\ $q=1.43$} & \makecell{$\pm$ \\ $\pm$} & \makecell{$4.14 \times 10^{-4}$ \\ $10^{-1}$} &
\makecell{$C=5.81 \times 10^{-16}$ \\ $q=6.67$} & \makecell{$\pm$ \\ $\pm$} & \makecell{$3.92 \times 10^{-16}$ \\ $1.68 \times 10^{-1}$} \\
\hline
\end{tabular}
\end{center}
 \begin{quote}
  \caption{\protect\footnotesize{Parameters of the fits in Figure~\ref{fig:npart_vs_alpha}. Note that the nomenclature for the parameters has been chosen to be the same as that of \citet{Nozawa2016}, who expresses the power law as $n(a) = Ca^{-q}$; see \citet{Nozawa2016} for further details.}} \label{tab:fit_parameters}
 \end{quote}
\end{table*}

The left side of the distribution, with the smallest grains formed at very low densities in the late stages of the dust formation (see, e.g.,  Figure~\ref{fig:alpha_vs_radius}), does not display any particular trend. The gas where these grains are located achieves homologous expansion after dust is formed, i.e., after $\simeq 5000$~days from the beginning of the simulation. We highlight that, even though the number of dust grains in this part of the distribution is large, the mass of the gas where they reside, and hence the dust mass, is quite low. Its mass corresponds to $\simeq 0.44 \%$ and $\simeq 1.32 \%$ of the ejecta mass for C and MgSiO$_3$, respectively. This is in line with the results obtained by \citet{Iaconi2019}, who showed that most of the ejecta has achieved homologous expansion by the $\simeq 5000$~days mark. 

The middle part of the distributions contain the majority of the ejecta mass, $\simeq 67 \%$ in the case of C grains and $61 \%$ in the case of MgSiO$_3$ grains. The power law slopes representing the distribution are $-1.62$ and $-1.43$, for C and MgSiO$_3$, respectively. All of these dust grains form in areas where the gas has already achieved homologous expansion. We highlight here that we performed the fit in logarithmic space, i.e., on the distribution obtained as $n(a) = dn / d \log{a}$, where $n$ is the number of dust grains.
As a result the slopes of the power law fits decrease by 1 if the distribution is fitted in linear space\footnote{Let us define the distribution derived in linear space as $\frac{dn}{da}$. This distribution and the one derived in logarithmic space are connected by the relation $\frac{dn}{da} = \frac{1}{a \ln{10}} \frac{dn}{d \log{a}}$. If the distribution derived in logarithmic space is a power law of the type $Ca^{-q}$, we obtain $\frac{dn}{da} = \frac{Ca^{-(q+1)}}{\ln{10}}$.}, resulting in exponents of $-2.62$ and $-2.43$. The linear slopes we obtain differ from the typical values estimated for interstellar dust grains, which are in the range $-3.5 \lesssim -q \lesssim -4$ (e.g., \citealt{Nozawa2013b}).

The right part of the distributions corresponds to the cusps pointing upward visible in Figure~\ref{fig:alpha_vs_radius} and also in this case the dust grains form in regions where the gas has achieved homologous expansion. The cusps are located between $\simeq 300$ and $\simeq 600$~AU for both C and MgSiO$_3$ and contain, respectively, $\simeq 32 \%$ and $\simeq 37 \%$ of the ejecta mass. This part of the distributions are fitted by power laws whose slopes are steeper than for smaller grains with exponents of $-6.56$ and $-6.67$, for C and MgSiO$_3$, respectively. These exponents correspond to $-7.56$ and $-7.67$ for a linear fit. Also in this case the linear slopes we obtain differ from the typical values estimated for interstellar dust grains. The presence of a steeper power law at larger grain size is a feature which appears also in other dust formation environments. For example, \citet{Nozawa2003} observe a crossover point between smaller and larger grain sizes, with a steeper slope for larger grain sizes, for dust formed in supernova ejecta. As mentioned in Section~\ref{ssec:locations_of_the_grains}, the radius of newly formed grains is mainly determined by the gas density at the onset of dust formation; larger dust grains form in denser gas, and vice versa. Hence, the grain size distributions given here reflect more or less the density distribution of SPH particles. This implies that different dynamical models would result in different power-law indices.

We also overplot the distribution for the outer dust in the case of C in cyan (Figure~\ref{fig:npart_vs_alpha}, panel (a)) and in the case of MgSiO$_3$ in pink (Figure~\ref{fig:npart_vs_alpha}, panel (b)), to those of the distribution at the end of the simulation for comparison. The inner dust population dominates the grain size distribution at the end of the simulation. This is particularly true for large grain sizes. Larger grains are in fact mostly formed in the inner dust population, so the outer population misses that part of the statistic.

\subsection{The shape of the dusty ejecta}
\label{ssec:shape_of_the_dusty_ejecta}
\begin{figure*}
\centering     
\subfigure[]{\includegraphics[scale=0.45, trim=0.0cm 0.0cm 0.0cm 0.0cm, clip]{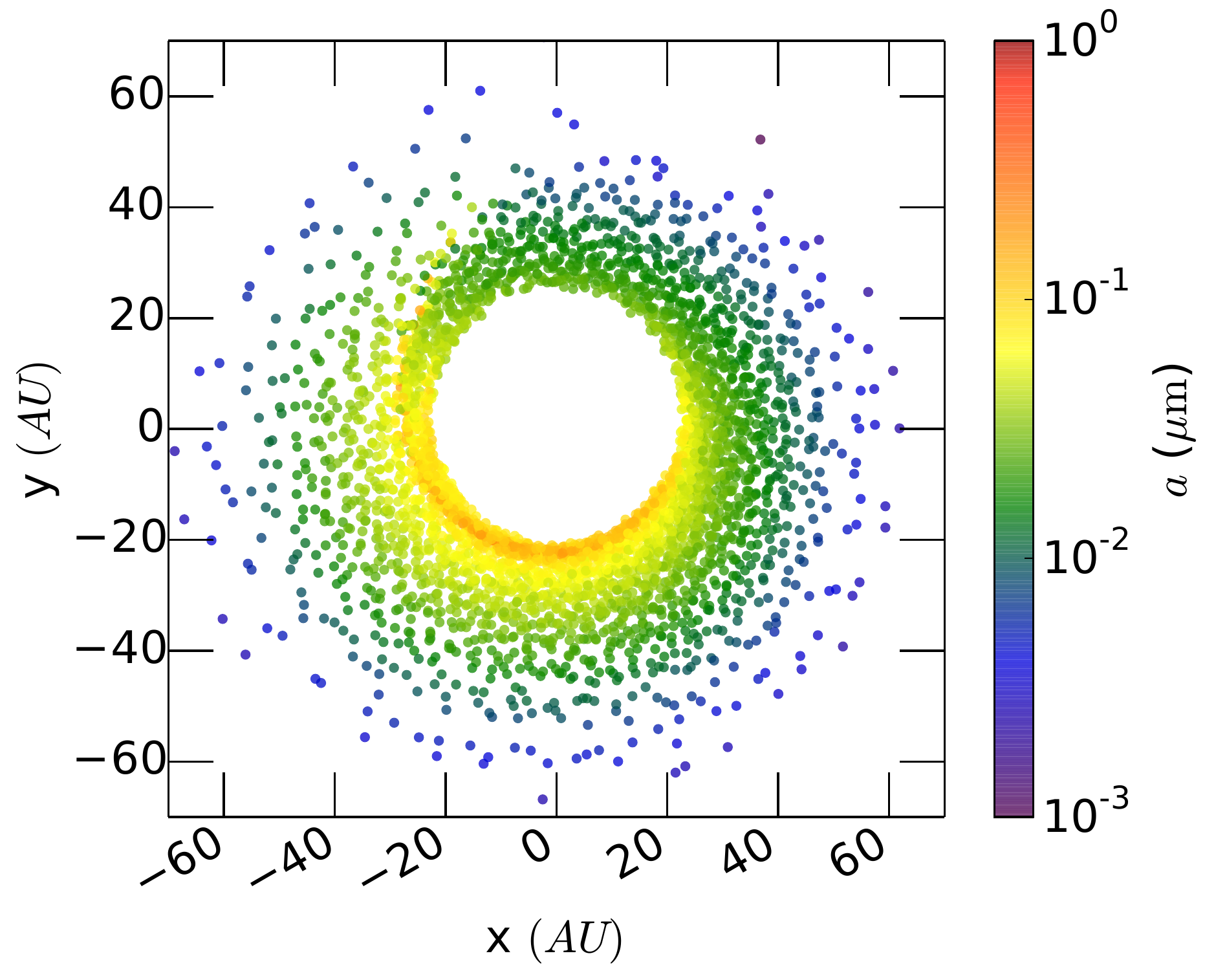}}
\subfigure[]{\includegraphics[scale=0.45, trim=0.0cm 0.0cm 0.0cm 0.0cm, clip]{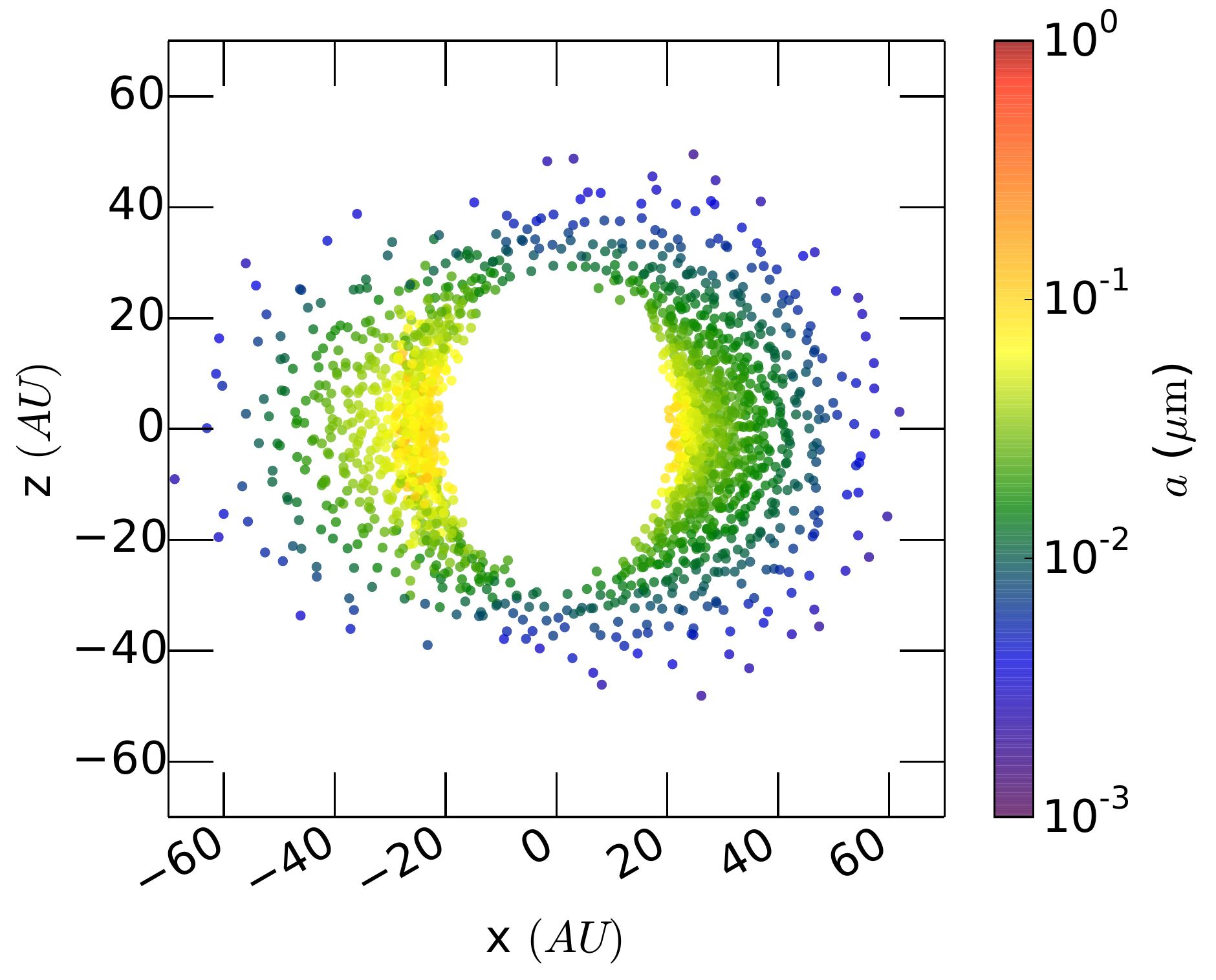}}
\subfigure[]{\includegraphics[scale=0.45, trim=0.0cm 0.0cm 0.0cm 0.0cm, clip]{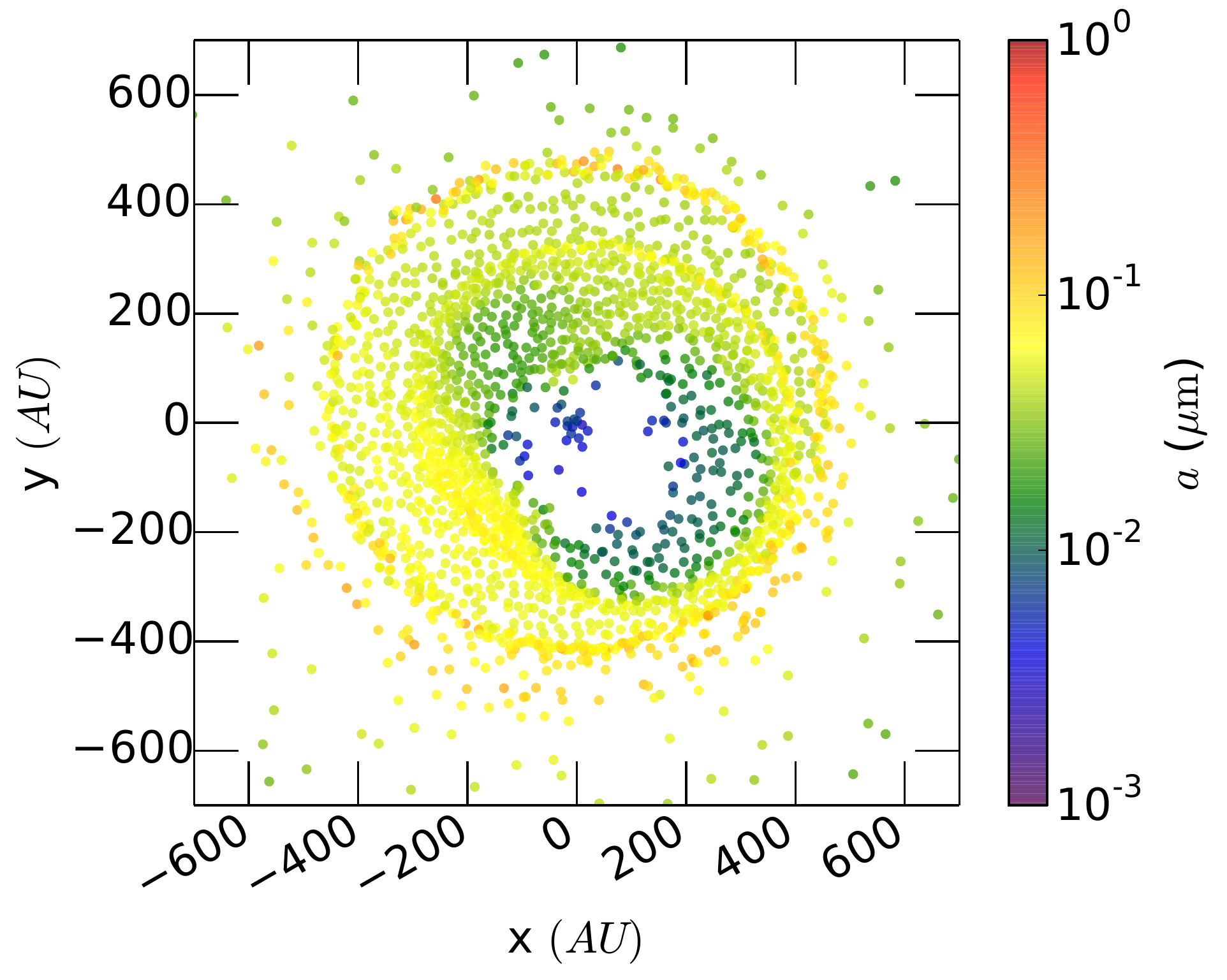}}
\subfigure[]{\includegraphics[scale=0.45, trim=0.0cm 0.0cm 0.0cm 0.0cm, clip]{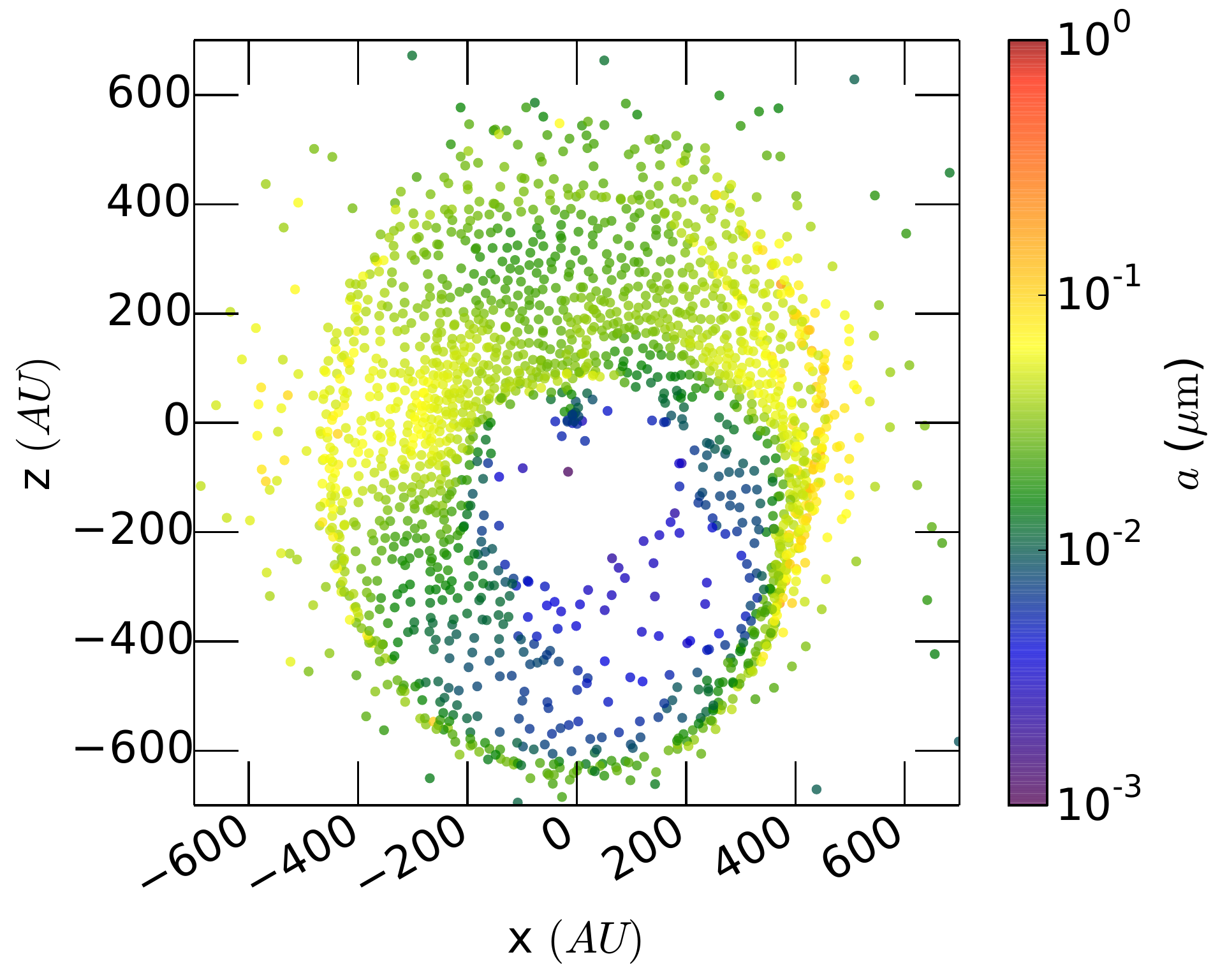}}
\caption{\protect\footnotesize{Slices on the $x-y$ (panels (a) and (c)) and $x-z$ (panels (b) and (d)) planes of the SPH particles that formed C grains, the dust size is represented by the colour scale. The slices are taken at $\simeq 900$~days (top) and at 18\,434~days (bottom), to clearly show the location of the two dust populations discussed in Section~\ref{ssec:locations_of_the_grains}. Note that panels (c) and (d) are cropped at 700~AU to include only the SPH particles belonging to the inner dust population, formed in the inner regions of the ejecta. We do not plot the map of the MgSiO$_3$ grains because is very similar to the one of the C grains, only with different grain sizes.}}
\label{fig:dust_map}
\end{figure*}

The two dust populations discussed in Section~\ref{ssec:locations_of_the_grains} differ mainly because the former, exterior one is more ordered, with smaller grains residing at larger distances. The latter population is instead more chaotic, with a range of grain sizes at all distances.
This is visible in the shape of the dusty ejecta, which we show in Figure~\ref{fig:dust_map} for the C grains, where the top row are slices on the $x-y$ and $x-z$ planes at $\simeq 900$~days and the bottom row are the same slices at the end of the simulation. We have chosen 900~days and 18\,434~days because they highlight well the separation between the two dust populations.

At $\simeq 900$~days we observe that the dusty part of the ejecta is only the external one, leaving a dust-free cavity around the central binary. Such cavity has an elliptic shape and spans an area between 20 and 30~AU just before the inner population starts forming and filling the dust-free space. In agreement with Figure~\ref{fig:alpha_vs_radius}, from the outside to the inside, we observe a trend of increasing grain size as a function of the distance. In the innermost area of the dusty gas there is a layer where the largest grains form (clearly visible as a mostly orange coloured area in panels a and b of Figure~\ref{fig:dust_map}). This results from the compression generated by the first orbit of the companion inside the primary's envelope. When the gas belonging to the compressed layer cools down to dust formation temperatures, its higher density prompts the formation of larger grains.
Moreover, the dust we observe in panels (a) and (b) of Figure~\ref{fig:dust_map} is mostly made up of unbound gas as it forms in the portion of the ejecta that is accelerated above escape velocity during the first orbit of the companion (see also Section~\ref{ssec:locations_of_the_grains}).

At the end of the simulation, the gas has rarefied and cooled enough to form dust over the entire ejecta. In panels (c) and (d) of Figure~\ref{fig:dust_map} we crop the slices to a distance of 700~AU, roughly corresponding to the transition distance between the outer and inner dust populations. In panel (c) we observe that the outermost area shows a greater number of larger grains, these are concentrated on the equatorial plane and correspond to those in the cusp between 400 and 600~AU in Figure~\ref{fig:alpha_vs_radius}. We also observe the presence of small grains in the most internal region of the ejecta. The remaining portion of the ejecta is occupied by a mixture of different grain sizes. There is a ring-shaped feature between $\simeq 300$ and $\simeq 400$~AU showing larger grain sizes. Since dust has formed over the entire ejecta, the patterns of the dusty ejecta are the same of those of the density distribution, which presents a higher density ring as shown in figures~4 and 5 of \citealt{Iaconi2019}.
A similar behaviour is present in the simulation of \citet{Reichardt2020}, who carried out a common envelope simulation that did not include the effects of dust but which, by virtue of using a tabulated equation of state, could map the opacity in the envelope as a function of time. In their figure 10, top panels, we see a thick high opacity ring at $\sim$100~AU from the centre. This ring compares directly to the ring observed in the outer population dust in Figure~\ref{fig:dust_map}, which sits at $\sim$30~AU. When we account for the fact that the snapshot presented by \citet{Reichardt2020} are taken at 1840~days, while those in Figure~\ref{fig:dust_map} are taken at 900~days (the outward velocity of the early ejecta is approximately 100~km~s$^{-1}$, justifying the different ring sizes). Finally, the dust distribution in the two perpendicular cuts also matches what is seen in the opacity maps of \citet{Reichardt2020}. Equatorially concentrated dust formation will propel different layers of the ejecta at different speeds contributing to their shaping in addition to what the geometry of the common envelope interaction has already imparted to the ejecta. 

\subsection{Dust mass}
\label{ssec:dust_mass}
\begin{figure*}
\centering     
\includegraphics[scale=0.45, trim=0.0cm 0.0cm 0.0cm 0.0cm, clip]{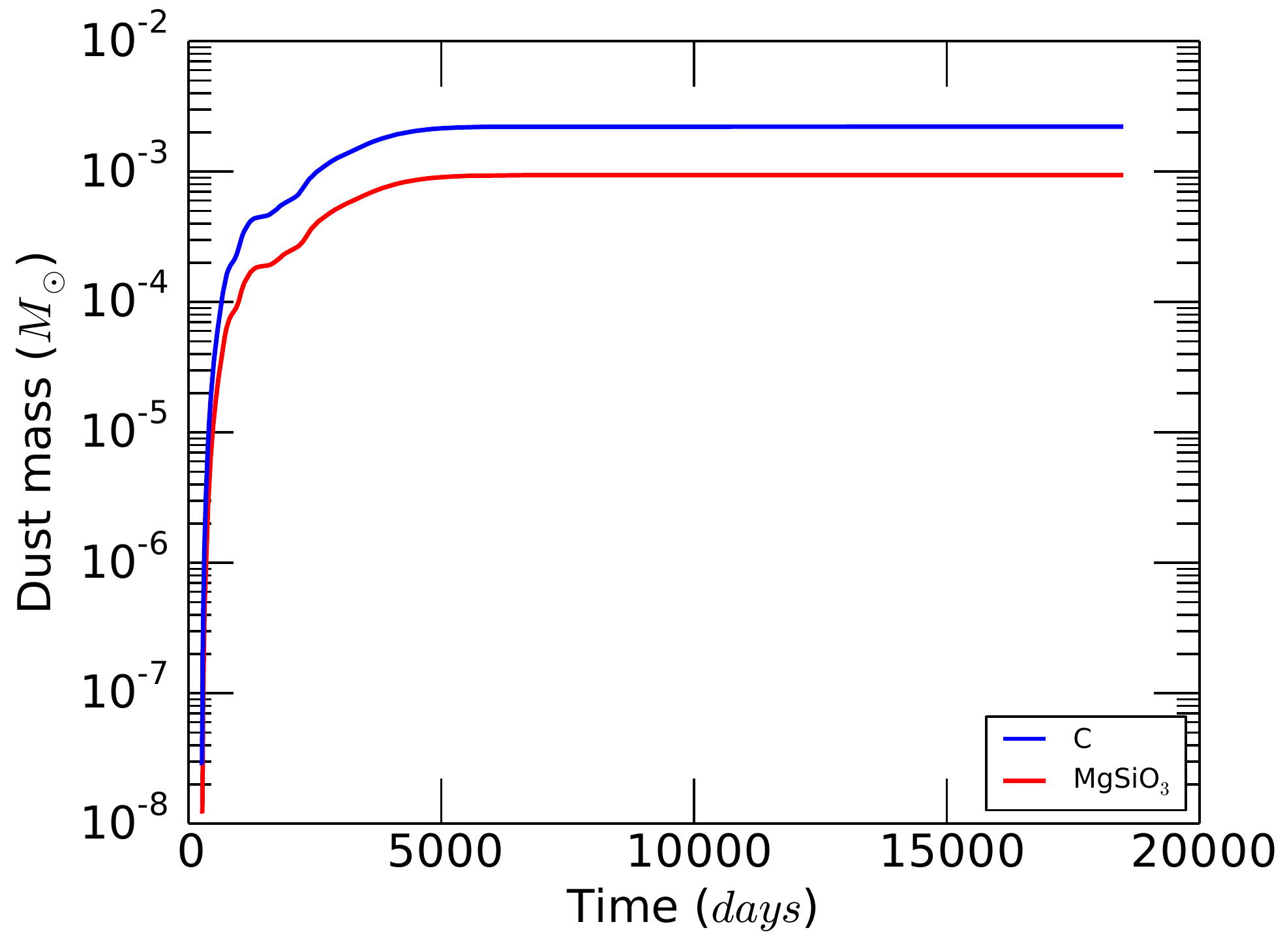}
\caption{\protect\footnotesize{Dust mass formed as a function of time for MgSiO$_3$ and C grains.}}
\label{fig:dust_mass_vs_time}
\end{figure*}

In our simulation every particle forms the same mass of either C or MgSiO$_3$ grains. This is because the amount of dust formed only depends on the key element number density and on the total mass of the SPH particle, with all the SPH particles in the simulation having the same mass. We show the evolution of the dust mass formed in Figure~\ref{fig:dust_mass_vs_time}. 

The mass of C grains at the end of the simulation amounts to $2.22\times10^{-3}$~\ms, while that of MgSiO$_3$ amounts to $9.39\times10^{-4}$~\ms. These values correspond to $\simeq0.5$~percent and $\simeq0.2$~percent of the ejected envelope mass, respectively. These values are in the same range to those estimated by \citet{Lu2013}, who calculated dust masses ranging between $\simeq 10^{-4}$~\ms \ and $5 \times 10^{-2}$~\ms \ for the bulk of their models. In the models of \citet{Reichardt2020}, approximately 0.06~\ms\ of gas reside in the high opacity shell where we presume dust is forming, implying a dust mass of $\sim6\times10^{-4}$\ms, using a gas-to-dust ratio of 100, a value that is in line with what we find here.

In Figure~\ref{fig:dust_mass_vs_time}, the shape of the curves for C and MgSiO$_3$ is similar. This is due to the fact that the supersaturation ratio reaches its critical value, $S = 10$, at the same time for both dust types, and when this happens the vast majority of the SPH particles have $\Lambda_{\rm on} > 1$, forming the maximum amount of dust possible (see Section~\ref{sec:dust_formation_process}). As a result, the timing of the production of dust is similar. 

The dust mass increases monotonically up to $\simeq 5000$~days, so new grains are constantly formed. As discussed previously in Sections~\ref{ssec:average_grain_size} and \ref{ssec:locations_of_the_grains}, we observe the layers of the ejecta gradually forming dust as they expand and cool down. From this it follows straight away that the total dust mass is steadily increasing.

\subsection{Contribution of CE to cosmic dust}
\label{ssec:contribution_of_CE_to_cosmic_dust}
In this section we estimate the CE contribution to the Galactic dust budget. To do so we need to estimate the number of CE events per year and multiply it by the dust yield of each as determined in Section~\ref{ssec:dust_mass} ($\simeq 10^{-3}$~\ms \ for both C and MgSiO$_3$ grains).


We can estimate the number of CE events per year for the Milky Way by assuming that they are the same as the rate of luminous red novae, namely $0.1 - 0.2$ events per year (\citealt{Kochanek2014}, \citealt{Howitt2020}). We note that this might be an overestimate, because we do not know what percentage of red novae is indeed the result of a CE event. 
With these numbers we obtain an injection of dust in the ISM by CE events of $1-2 \times 10^{-4}$~\ms~yr$^{-1}$. The dust injected in the ISM by sources other than supernovae or novae, namely all stellar winds, is estimated to be $\sim 4.5 \times 10^{-3}$~\ms~yr$^{-1}$ for the Milky Way (\citealt{Draine2009}). The CE contribution to the Galactic dust can therefore be estimated to be up to $5 \%$ of the dust produced by other known sources. The estimate provided here gives an idea of the fact that the injection of dust into the ISM by CE does not dominate the cosmic dust production, but might not be negligible and could be comparable to other dust sources such as novae and SNe (see, e.g., table~1 of \citealt{Draine2009}).

\citet{Lu2013} also considered the Galactic dust production from CE events over the Galaxy's lifetime using an analytical estimate of the dust production and a population synthesis code to determine the rate of CE occurrences. They use a dust yield per event similar to ours, and an estimated $20 \%$ of all systems that go through a CE interaction, which corresponds to a rate up to $0.8$ events per year. As a result of their four times higher frequency of CE interactions, they obtain a larger yield of dust by CE events.  


\section{Observational counterparts}
\label{sec:observational_counterparts}
The presence of dust is a very important feature frequently identified in the observed counterparts of CE events (see Section~\ref{sec:introduction}). However, the detailed study of the properties of the newly formed dust grains has been limited to the CE merger V1309 Sco, between a solar mass subgiant star and a much less massive companion (\citealt{Nicholls2013} and \citealt{Tylenda2016}).

\citet{Nicholls2013} analysed the dust formation between $\simeq 547$~days and $\simeq 700$~days after the outburst by constructing a simple dust model based on their data. Their model considers a couple of dust species, grain sizes between $0.1$ and $4.0$~$\mathrm{\mu m}$ and temperatures between $200$ and $1500$~K. Within these ranges they selected the dust type and size that best fit the observational data. They concluded that the continuum was well represented by amorphous silicates dust at $400$~K, while the main absorption line could be fitted by amorphous pyroxene dust of $3$~$\mathrm{\mu m}$ at $800$~K. The warm temperature and large grain size indicated by the line feature led them to conclude that the dust had recently formed and was processed in the circumstellar environment, rather than inside the stellar envelope, thus confirming its CE origin.

\citet{Tylenda2016} created instead a model aimed to determine general parameters of the dust ejected and study its interaction with the central object, rather than to determine the chemical composition of the dust. Their main assumptions were therefore that the dust was composed of silicates and that its spatial distribution was that of a shell cut in the polar direction, so to simulate dust concentrated around the orbital plane. Their radiation transfer calculation allowed them to estimate a total dust mass of $\simeq 10^{-3}$~\ms, which reproduced well the observed spectrum. They also noted that the observations showed evidence of light penetrating the ejecta in certain directions without interacting significantly with the dust.

Our dust formation model is based on a system where the primary has a mass similar to the primary star of V1309 Sco, but is substantially larger, and our companion is $\sim$4 times more massive (for a range of possible parameters of this system see e.g., \citealt{Tylenda2011}). Despite these differences we carry out a comparison.

The V1309 Sco merger took place during the very early RGB phase of the primary, so we will be looking at the MgSiO$_3$ grains for a comparison, a silicate mineral such as the one contemplated by \citet{Nicholls2013}. According to our model, dust would form between $\simeq 300$ and $\simeq 5000$~days from the beginning of the simulation, where $\simeq 300$~days marks the end of the dynamic in-spiral phase. During this period both temperature and density decrease by one order of magnitude in the inner parts of the ejecta to several orders of magnitude in the outskirts. We observe that dust forms as soon as the gas decreases to temperatures  $\sim$1000~K. Therefore we predict that dust forms in warm environments, similar to the conclusion of \citet{Nicholls2013}. On the other hand, the entire MgSiO$_3$ dust population in our simulation has sizes smaller than $0.1$~$\mathrm{\mu m}$. It is therefore possible that in our case dust forms at densities low enough to counter balance the warm temperatures and still produce small grains. 

Given the differences and the approximations in our model and in the observation-derived quantities it is impossible to perform an accurate quantitative comparison. However, we can make a few points:
\vspace{-0.2cm}
\begin{enumerate}
    \item The formation of large dust particles usually requires extreme environments and/or special physical processes (e.g., coagulation via grain-grain collisions). In the current calculation we do not include such processes and therefore, even if a population of larger grains could be formed, we would not be able to reproduce it. Nevertheless, given the typical densities and temperatures of our gas, it seems unlikely that larger grains would be able to form.
    \item Grain sizes derived from IR-SED fitting tend to be large ($\gtrsim 1$~$\mathrm{\mu m}$). This is known for the case of nova ejecta, type II SN, AGB stars and PN (\citealt{Meixner2002}, \citealt{Wesson2010}, \citealt{Sakon2016}). Therefore, the study of \citet{Nicholls2013} might have obtained systematically large grain sizes for V1309 Sco.
    \item \citet{Nicholls2013} concluded that the dust was formed in the post dynamic in-spiral phase. However, there is another possible scenario for the dust formation in V1309 Sco. As can be seen in figure~1 of \citet{Nicholls2013}, the optical light curve quickly declines during the year 2007, roughly one year before the outburst. In the phase before the outburst, dust has formed in gas lost from the binary prior to the dynamic in-spiral (\citealt{Tylenda2016}). These dust grains could then have been swept up by the material ejected during the dynamic in-spiral and heated up to temperatures high enough to emit in the IR band. In such a scenario, dust can form in physical conditions more suitable for the formations of grains with sizes $\gtrsim 1$~$\mathrm{\mu m}$.
    \item The mass of the dust we obtain is similar to that of \citet{Tylenda2016}, this gives a good indication that the dust formed after the dynamic in-spiral is stable and does not get destroyed during the subsequent expansion of the ejecta (see also our argument on shocks in Section~\ref{ssec:application_of_the_model_to_the_code_data}).
    \item the distribution of the dust we obtain reflects the overall geometry of the ejecta. This not only leaves a more rarefied layer of dust towards the polar direction, because the CE ejecta are naturally concentrated along the equatorial plane, but also leaves thinner layers of dust in certain directions along the equatorial plane itself (see for example the bottom right section of the dusty ejecta in panel(c) of Figure~\ref{fig:dust_map}). This could allow the radiation from the central object to escape from these spots, similarly to what was observed by \citet{Tylenda2016}. 
\end{enumerate}

A more adequate comparison would require more observations of dust in post-CE systems together with more theoretical dust formation models based on CE simulations. These are both missing in the current literature.


\section{Summary and conclusions}
\label{sec:summary_and_conclusions}
In this paper, we presented the first work on dust formation based on a 3D hydrodynamic simulation of the CE interaction. We extended the work carried out by \citet{Iaconi2019} by applying the dust formation model of \citet{Nozawa2013} to the simulation outputs of the 3D SPH hydrodynamic simulation of the CE binary interaction of \citet{Iaconi2019}. In this simulation, the system has been evolved for $18\,434$~days ($\simeq 50$~yr) after the termination of the dynamic in-spiral to assess the dynamic and thermodynamic behaviour of the ejecta. This allows us to investigate the details of the formation of dust, its spatial and size distributions and to estimate the contribution of the CE interaction to the production of cosmic dust. We consider two types of dust, carbon dust (C) and pyroxene-type silicates dust (MgSiO$_3$), that represent the most common types of grains present in the interstellar medium.
Our main results are the following:
\begin{enumerate}
    \item In both cases of C and MgSiO$_3$, dust grains are produced all over the ejected envelope (i.e., in $99 \%$ of the SPH particles). The dust formation starts right after the end of the dynamic in-spiral ($\simeq 300$~days) and terminates at $\simeq 5000$~days. 
    \item The dust formed can be divided into two populations, an outer one one formed between $\simeq 300$ and $\simeq 900$~days, and an inner one formed between $\simeq 900$ and $\simeq 5000$~days from the rest of the ejecta. In the outer population dust is formed in a more ordered way with respect to the inner population, starting at larger distances and with smaller grains and terminating at smaller distances with larger grains. The inner population is more complex, with multiple grain sizes forming in the same layers at different times. 
    The formation of two populations is the result of different envelope ejection dynamics. The outer population is formed in the gas that belongs to the outer layers of the primary, which are rapidly unbound during the first orbit of the companion inside the CE and expand with negligible self-interaction in the empty space around the binary. The inner population is formed in the gas that belongs to the inner layers of the primary; such layers interact with one another, achieving the conditions for dust formation under different combinations of physical parameters inside the same layer.
    \item On average, the size of dust grains formed increases as time passes. At the end of the dust formation process the average radius of the grains is $\simeq 4 \times 10^{-2}$~$\rm \mu m$ for C and $\simeq 5 \times 10^{-3}$~$\rm \mu m$ for MgSiO$_3$.
    \item For both C and MgSiO$_3$ the size distribution of the dust formed is weighted towards small grains and spans a relatively broad range. The distributions can be fitted with a double power law, with a turnover point where the slope of the fitting power law changes. The power law fitting the smaller grain sizes is flatter than the one for the larger grain sizes, a feature present also in other astrophysical environments such as supernovae (\citealt{Nozawa2003}). However, the indeces of the two power laws differ from the canonical value expected for the interstellar dust grains. To clarify what is happening, the same dust formation calculation should be carried out for several CE simulations. This would allow us to see, e.g., whether this is a common feature of different CE systems.
    \item The mass of the dust formed in the ejecta is $\simeq 10^{-3}$~\ms \ for both C and MgSiO$_3$. By assuming this value as a standard dust yield from CE interactions we estimated the contribution of CE to cosmic dust to be  $1-2 \times 10^{-4}$~\ms~yr$^{-1}$, or $\simeq 5 \%$ of that produced by main dust sources, i.e., stellar winds. Such a contribution is not negligible, and comparable to that of novae and SNe. 
    \item A comparison with an observational study of dust formation in the post-CE merger V1309 Sco by \citet{Nicholls2013} shows that we have quite different results in terms of the grain sizes obtained, with our grains being much smaller than those estimated from the observations. However, there are several caveats in both our theoretical work and the procedure used by \citet{Nicholls2013}, which make a comparison difficult. Our work shows instead agreement in terms of dust mass and spatial distribution with the results of \citet{Tylenda2016}, who also modelled V1309 Sco.
\end{enumerate}

This is the first work on dust formation based on a 3D hydrodynamic simulation of the CE interaction. We hope that in the future more work of this kind will be performed with different stellar masses, orbital parameters, dust formation models and numerical codes to achieve a deeper understanding of the dust formation process during the CE evolution. It would be important to also assess the dynamic effect that the presence of dust has on the ejecta.


\section*{Acknowledgments}
\label{sec:acknowledgments}
RI is grateful for the financial support provided by the Postodoctoral Research Fellowship of the Japan Society for the Promotion of Science (JSPS P18753). KM acknowledges the support provided by the JSPS KAKENHI grant 17H02864, 18H04585, 18H05223, 20H00174 and 20H04737. TN acknowledges the support provided by the JSPS KAKENHI grant 18K03707.
We are grateful to the referee for the useful comments and for helping to improve the paper. We also thank Noam Soker and Tomasz Kaminski for their feedback after posting this work on astro-ph, which significantly contributed to deepening the arguments proposed in this work.
We acknowledge the computational facilities of the Department of Physics and Astronomy of Macquarie University, on which the simulations have been carried out. The computations described in this work were performed using the PHANTOM code (https://phantomsph.bitbucket.io/), which is the product of a collaborative effort of scientists under the lead of A/Prof. Daniel Price (Monash University, Melbourne, VIC, Australia).


\section*{Data availability}
\label{sec:data_availability}
All simulation outputs are available upon request by e-mailing roberto.iaconi@kusastro.kyoto-u.ac.jp.


\bibliographystyle{aa}
\bibliography{bibliography}{}

\begin{thebibliography}{67}
\expandafter\ifx\csname natexlab\endcsname\relax\def\natexlab#1{#1}\fi

\bibitem[{{Asplund} {et~al.}(2009){Asplund}, {Grevesse}, {Sauval}, \&
  {Scott}}]{Asplund2009}
{Asplund}, M., {Grevesse}, N., {Sauval}, A.~J., \& {Scott}, P. 2009, Annual
  Review of Astronomy and Astrophysics, 47, 481

\bibitem[{{Banerjee} {et~al.}(2007){Banerjee}, {Misselt}, {Su}, {Ashok}, \&
  {Smith}}]{Banerjee2007}
{Banerjee}, D.~P.~K., {Misselt}, K.~A., {Su}, K.~Y.~L., {Ashok}, N.~M., \&
  {Smith}, P.~S. 2007, \apjl, 666, L25

\bibitem[{{Banerjee} {et~al.}(2015){Banerjee}, {Nuth}, {Misselt}, {Varricatt},
  {Sand}, {Ashok}, {Su}, {Marion}, \& {Marengo}}]{Banerjee2015}
{Banerjee}, D. P.~K., {Nuth}, Joseph~A., I., {Misselt}, K.~A., {et~al.} 2015,
  \apj, 814, 109

\bibitem[{{Clayton} {et~al.}(1999){Clayton}, {Liu}, \&
  {Dalgarno}}]{Clayton1999}
{Clayton}, D.~D., {Liu}, W., \& {Dalgarno}, A. 1999, Science, 283, 1290

\bibitem[{{Clayton} {et~al.}(2017){Clayton}, {Podsiadlowski}, {Ivanova}, \&
  {Justham}}]{Clayton2017}
{Clayton}, M., {Podsiadlowski}, P., {Ivanova}, N., \& {Justham}, S. 2017,
  \mnras, 470, 1788

\bibitem[{{De} {et~al.}(2019){De}, {MacLeod}, {Everson}, {Antoni}, {Mandel}, \&
  {Ramirez-Ruiz}}]{De2019}
{De}, S., {MacLeod}, M., {Everson}, R.~W., {et~al.} 2019, arXiv e-prints,
  arXiv:1910.13333

\bibitem[{{Draine}(2009)}]{Draine2009}
{Draine}, B.~T. 2009, in Astronomical Society of the Pacific Conference Series,
  Vol. 414, Cosmic Dust - Near and Far, ed. T.~{Henning}, E.~{Gr{\"u}n}, \&
  J.~{Steinacker} (Astronomical Society of the Pacific), 453

\bibitem[{{Evans} {et~al.}(2005){Evans}, {Tyne}, {Smith}, {Geballe},
  {Rawlings}, \& {Eyres}}]{Evans2005}
{Evans}, A., {Tyne}, V.~H., {Smith}, O., {et~al.} 2005, \mnras, 360, 1483

\bibitem[{{Ferrarotti} \& {Gail}(2006)}]{Ferrarotti2006}
{Ferrarotti}, A.~S. \& {Gail}, H.~P. 2006, \aap, 447, 553

\bibitem[{{Glanz} \& {Perets}(2018)}]{Glanz2018}
{Glanz}, H. \& {Perets}, H.~B. 2018, \mnras, 478, L12

\bibitem[{{Grichener} {et~al.}(2018){Grichener}, {Sabach}, \&
  {Soker}}]{Grichener2018}
{Grichener}, A., {Sabach}, E., \& {Soker}, N. 2018, \mnras, 478, 1818

\bibitem[{{Hardy} {et~al.}(2016){Hardy}, {Schreiber}, {Parsons}, {Caceres},
  {Brinkworth}, {Veras}, {G{\"a}nsicke}, {Marsh}, \& {Cieza}}]{Hardy2016}
{Hardy}, A., {Schreiber}, M.~R., {Parsons}, S.~G., {et~al.} 2016, \mnras, 459,
  4518

\bibitem[{{Howitt} {et~al.}(2020){Howitt}, {Stevenson}, {Vigna-G{\'o}mez},
  {Justham}, {Ivanova}, {Woods}, {Neijssel}, \& {Mandel}}]{Howitt2020}
{Howitt}, G., {Stevenson}, S., {Vigna-G{\'o}mez}, A.~r., {et~al.} 2020, \mnras,
  30

\bibitem[{{Iaconi} \& {De Marco}(2019)}]{Iaconi2019b}
{Iaconi}, R. \& {De Marco}, O. 2019, \mnras, 490, 2550

\bibitem[{{Iaconi} {et~al.}(2018){Iaconi}, {De Marco}, {Passy}, \&
  {Staff}}]{Iaconi2018}
{Iaconi}, R., {De Marco}, O., {Passy}, J.-C., \& {Staff}, J. 2018, \mnras, 477,
  2349

\bibitem[{{Iaconi} {et~al.}(2019){Iaconi}, {Maeda}, {De Marco}, {Nozawa}, \&
  {Reichardt}}]{Iaconi2019}
{Iaconi}, R., {Maeda}, K., {De Marco}, O., {Nozawa}, T., \& {Reichardt}, T.
  2019, \mnras, 489, 3334

\bibitem[{{Iaconi} {et~al.}(2017){Iaconi}, {Reichardt}, {Staff}, {De Marco},
  {Passy}, {Price}, {Wurster}, \& {Herwig}}]{Iaconi2017}
{Iaconi}, R., {Reichardt}, T., {Staff}, J., {et~al.} 2017, \mnras, 464, 4028

\bibitem[{{Ivanova}(2018)}]{Ivanova2018}
{Ivanova}, N. 2018, \apj, 858, L24

\bibitem[{{Ivanova} {et~al.}(2013{\natexlab{a}}){Ivanova}, {Justham}, {Avendano
  Nandez}, \& {Lombardi}}]{Ivanova2013b}
{Ivanova}, N., {Justham}, S., {Avendano Nandez}, J.~L., \& {Lombardi}, J.~C.
  2013{\natexlab{a}}, Science, 339, 433

\bibitem[{{Ivanova} {et~al.}(2013{\natexlab{b}}){Ivanova}, {Justham}, {Chen},
  {De Marco}, {Fryer}, {Gaburov}, {Ge}, {Glebbeek}, {Han}, {Li}, {Lu}, {Marsh},
  {Podsiadlowski}, {Potter}, {Soker}, {Taam}, {Tauris}, {van den Heuvel}, \&
  {Webbink}}]{Ivanova2013}
{Ivanova}, N., {Justham}, S., {Chen}, X., {et~al.} 2013{\natexlab{b}}, \aapr,
  21, 59

\bibitem[{{Ivanova} \& {Nandez}(2016)}]{Ivanova2016}
{Ivanova}, N. \& {Nandez}, J.~L.~A. 2016, \mnras, 462, 362

\bibitem[{{Jencson} {et~al.}(2019){Jencson}, {Adams}, {Bond}, {van Dyk},
  {Kasliwal}, {Bally}, {Blagorodnova}, {De}, {Fremling}, {Yao}, {Fruchter},
  {Rubin}, {Barbarino}, {Sollerman}, {Miller}, {Hicks}, {Malkan}, {Andreoni},
  {Bellm}, {Buchheim}, {Dekany}, {Feeney}, {Frederick}, {Gal-Yam}, {Gehrz},
  {Giomi}, {Graham}, {Green}, {Hale}, {Hankins}, {Hanson}, {Helou}, {Ho},
  {Hung}, {Juri{\'c}}, {Kendurkar}, {Kulkarni}, {Lau}, {Masci}, {Neill},
  {Quin}, {Riddle}, {Rusholme}, {Sims}, {Smith}, {Smith}, {Soumagnac},
  {Tachibana}, {Tinyanont}, {Walters}, {Watson}, \& {Williams}}]{Jencson2019}
{Jencson}, J.~E., {Adams}, S.~M., {Bond}, H.~E., {et~al.} 2019, \apjl, 880, L20

\bibitem[{{Keith} \& {Lazzati}(2011)}]{Keith2011}
{Keith}, A.~C. \& {Lazzati}, D. 2011, \mnras, 410, 685

\bibitem[{{Kochanek} {et~al.}(2014){Kochanek}, {Adams}, \&
  {Belczynski}}]{Kochanek2014}
{Kochanek}, C.~S., {Adams}, S.~M., \& {Belczynski}, K. 2014, \mnras, 443, 1319

\bibitem[{{Kuruwita} {et~al.}(2016){Kuruwita}, {Staff}, \& {De
  Marco}}]{Kuruwita2016}
{Kuruwita}, R.~L., {Staff}, J., \& {De Marco}, O. 2016, \mnras, 461, 486

\bibitem[{{L{\'o}pez-C{\'a}mara} {et~al.}(2019){L{\'o}pez-C{\'a}mara}, {De
  Colle}, \& {Moreno M{\'e}ndez}}]{Lopez-camara2019}
{L{\'o}pez-C{\'a}mara}, D., {De Colle}, F., \& {Moreno M{\'e}ndez}, E. 2019,
  \mnras, 482, 3646

\bibitem[{{L{\"u}} {et~al.}(2013){L{\"u}}, {Zhu}, \& {Podsiadlowski}}]{Lu2013}
{L{\"u}}, G., {Zhu}, C., \& {Podsiadlowski}, P. 2013, \apj, 768, 193

\bibitem[{{MacLeod} {et~al.}(2017){MacLeod}, {Antoni}, {Murguia-Berthier},
  {Macias}, \& {Ramirez-Ruiz}}]{MacLeod2017b}
{MacLeod}, M., {Antoni}, A., {Murguia-Berthier}, A., {Macias}, P., \&
  {Ramirez-Ruiz}, E. 2017, \apj, 838, 56

\bibitem[{{Meixner} {et~al.}(2002){Meixner}, {Ueta}, {Bobrowsky}, \&
  {Speck}}]{Meixner2002}
{Meixner}, M., {Ueta}, T., {Bobrowsky}, M., \& {Speck}, A. 2002, \apj, 571, 936

\bibitem[{{Murguia-Berthier} {et~al.}(2017){Murguia-Berthier}, {MacLeod},
  {Ramirez-Ruiz}, {Antoni}, \& {Macias}}]{Murguia-Berthier2017}
{Murguia-Berthier}, A., {MacLeod}, M., {Ramirez-Ruiz}, E., {Antoni}, A., \&
  {Macias}, P. 2017, \apj, 845, 173

\bibitem[{{Nandez} \& {Ivanova}(2016)}]{Nandez2016}
{Nandez}, J.~L.~A. \& {Ivanova}, N. 2016, \mnras, 460, 3992

\bibitem[{{Nandez} {et~al.}(2015){Nandez}, {Ivanova}, \&
  {Lombardi}}]{Nandez2015}
{Nandez}, J.~L.~A., {Ivanova}, N., \& {Lombardi}, J.~C. 2015, \mnras, 450, L39

\bibitem[{{Nicholls} {et~al.}(2013){Nicholls}, {Melis}, {Soszynski}, {Udalski},
  {Szymanski}, {Kubiak}, {Pietrzynski}, {Poleski}, {Ulaczyk}, \&
  {Wyrzykowski}}]{Nicholls2013}
{Nicholls}, C.~P., {Melis}, C., {Soszynski}, I., {et~al.} 2013, \mnras, 431,
  L33

\bibitem[{{Nozawa}(2016)}]{Nozawa2016}
{Nozawa}, T. 2016, \planss, 133, 36

\bibitem[{{Nozawa} \& {Fukugita}(2013)}]{Nozawa2013b}
{Nozawa}, T. \& {Fukugita}, M. 2013, \apj, 770, 27

\bibitem[{{Nozawa} \& {Kozasa}(2013)}]{Nozawa2013}
{Nozawa}, T. \& {Kozasa}, T. 2013, \apj, 776, 24

\bibitem[{{Nozawa} {et~al.}(2006){Nozawa}, {Kozasa}, \& {Habe}}]{Nozawa2006}
{Nozawa}, T., {Kozasa}, T., \& {Habe}, A. 2006, \apj, 648, 435

\bibitem[{{Nozawa} {et~al.}(2007){Nozawa}, {Kozasa}, {Habe}, {Dwek}, {Umeda},
  {Tominaga}, {Maeda}, \& {Nomoto}}]{Nozawa2007}
{Nozawa}, T., {Kozasa}, T., {Habe}, A., {et~al.} 2007, \apj, 666, 955

\bibitem[{{Nozawa} {et~al.}(2008){Nozawa}, {Kozasa}, {Tominaga}, {Sakon},
  {Tanaka}, {Suzuki}, {Nomoto}, {Maeda}, {Umeda}, {Limongi}, \&
  {Onaka}}]{Nozawa2008}
{Nozawa}, T., {Kozasa}, T., {Tominaga}, N., {et~al.} 2008, \apj, 684, 1343

\bibitem[{{Nozawa} {et~al.}(2003){Nozawa}, {Kozasa}, {Umeda}, {Maeda}, \&
  {Nomoto}}]{Nozawa2003}
{Nozawa}, T., {Kozasa}, T., {Umeda}, H., {Maeda}, K., \& {Nomoto}, K. 2003,
  \apj, 598, 785

\bibitem[{{Nozawa} {et~al.}(2014){Nozawa}, {Yoon}, {Maeda}, {Kozasa}, {Nomoto},
  \& {Langer}}]{Nozawa2014}
{Nozawa}, T., {Yoon}, S.-C., {Maeda}, K., {et~al.} 2014, \apjl, 787, L17

\bibitem[{{Ohlmann} {et~al.}(2016{\natexlab{a}}){Ohlmann}, {R{\"o}pke},
  {Pakmor}, \& {Springel}}]{Ohlmann2016}
{Ohlmann}, S.~T., {R{\"o}pke}, F.~K., {Pakmor}, R., \& {Springel}, V.
  2016{\natexlab{a}}, \apjl, 816, L9

\bibitem[{{Ohlmann} {et~al.}(2016{\natexlab{b}}){Ohlmann}, {R{\"o}pke},
  {Pakmor}, {Springel}, \& {M{\"u}ller}}]{Ohlmann2016b}
{Ohlmann}, S.~T., {R{\"o}pke}, F.~K., {Pakmor}, R., {Springel}, V., \&
  {M{\"u}ller}, E. 2016{\natexlab{b}}, \mnras, 462, L121

\bibitem[{{Paczynski}(1976)}]{Paczynski1976}
{Paczynski}, B. 1976, in IAU Symposium, Vol.~73, Structure and Evolution of
  Close Binary Systems, ed. P.~{Eggleton}, S.~{Mitton}, \& J.~{Whelan}, 75

\bibitem[{{Paquette} \& {Nuth}(2011)}]{Paquette2011}
{Paquette}, J.~A. \& {Nuth}, Joseph~A., I. 2011, \apjl, 737, L6

\bibitem[{{Passy} {et~al.}(2012){Passy}, {De Marco}, {Fryer}, {Herwig},
  {Diehl}, {Oishi}, {Mac Low}, {Bryan}, \& {Rockefeller}}]{Passy2012}
{Passy}, J.-C., {De Marco}, O., {Fryer}, C.~L., {et~al.} 2012, \apj, 744, 52

\bibitem[{{Pastorello} {et~al.}(2019){Pastorello}, {Chen}, {Cai},
  {Morales-Garoffolo}, {Cano}, {Mason}, {Barsukova}, {Benetti}, {Berton}, \&
  {Bose}}]{Pastorello2019}
{Pastorello}, A., {Chen}, T.~W., {Cai}, Y.~Z., {et~al.} 2019, \aap, 625, L8

\bibitem[{{Paxton} {et~al.}(2011){Paxton}, {Bildsten}, {Dotter}, {Herwig},
  {Lesaffre}, \& {Timmes}}]{Paxton2011}
{Paxton}, B., {Bildsten}, L., {Dotter}, A., {et~al.} 2011, \apjs, 192, 3

\bibitem[{{Reichardt} {et~al.}(2020){Reichardt}, {De Marco}, {Iaconi},
  {Chamandy}, \& {Price}}]{Reichardt2020}
{Reichardt}, T.~A., {De Marco}, O., {Iaconi}, R., {Chamandy}, L., \& {Price},
  D.~J. 2020, \mnras, 494, 5333

\bibitem[{{Reichardt} {et~al.}(2019){Reichardt}, {De Marco}, {Iaconi}, {Tout},
  \& {Price}}]{Reichardt2019}
{Reichardt}, T.~A., {De Marco}, O., {Iaconi}, R., {Tout}, C.~A., \& {Price},
  D.~J. 2019, \mnras, 484, 631

\bibitem[{{Ricker} \& {Taam}(2012)}]{Ricker2012}
{Ricker}, P.~M. \& {Taam}, R.~E. 2012, \apj, 746, 74

\bibitem[{{Sakon} {et~al.}(2016){Sakon}, {Sako}, {Onaka}, {Nozawa}, {Kimura},
  {Fujiyoshi}, {Shimonishi}, {Usui}, {Takahashi}, {Ohsawa}, {Arai}, {Uemura},
  {Nagayama}, {Koo}, \& {Kozasa}}]{Sakon2016}
{Sakon}, I., {Sako}, S., {Onaka}, T., {et~al.} 2016, \apj, 817, 145

\bibitem[{{Sandquist} {et~al.}(1998){Sandquist}, {Taam}, {Chen}, {Bodenheimer},
  \& {Burkert}}]{Sandquist1998}
{Sandquist}, E.~L., {Taam}, R.~E., {Chen}, X., {Bodenheimer}, P., \& {Burkert},
  A. 1998, \apj, 500, 909

\bibitem[{{Schreier} {et~al.}(2019){Schreier}, {Hillel}, \&
  {Soker}}]{Schreier2019}
{Schreier}, R., {Hillel}, S., \& {Soker}, N. 2019, \mnras, 490, 4748

\bibitem[{{Shiber} {et~al.}(2019){Shiber}, {Iaconi}, {De Marco}, \&
  {Soker}}]{Shiber2019}
{Shiber}, S., {Iaconi}, R., {De Marco}, O., \& {Soker}, N. 2019, \mnras, 488,
  5615

\bibitem[{{Shiber} {et~al.}(2017){Shiber}, {Kashi}, \& {Soker}}]{Shiber2017}
{Shiber}, S., {Kashi}, A., \& {Soker}, N. 2017, \mnras, 465, L54

\bibitem[{{Shiber} \& {Soker}(2018)}]{Shiber2018}
{Shiber}, S. \& {Soker}, N. 2018, \mnras, 477, 2584

\bibitem[{{Soker}(1993)}]{Soker1993}
{Soker}, N. 1993, \apj, 417, 347

\bibitem[{{Soker}(1998)}]{Soker1998}
{Soker}, N. 1998, \mnras, 299, 1242

\bibitem[{{Soker}(2017)}]{Soker2017}
{Soker}, N. 2017, \mnras, 471, 4839

\bibitem[{{Soker} {et~al.}(2018){Soker}, {Grichener}, \& {Sabach}}]{Soker2018}
{Soker}, N., {Grichener}, A., \& {Sabach}, E. 2018, \apj, 863, L14

\bibitem[{{Staff} {et~al.}(2016){Staff}, {De Marco}, {Macdonald}, {Galaviz},
  {Passy}, {Iaconi}, \& {Low}}]{Staff2016a}
{Staff}, J.~E., {De Marco}, O., {Macdonald}, D., {et~al.} 2016, \mnras, 455,
  3511

\bibitem[{{Tylenda} {et~al.}(2011){Tylenda}, {Hajduk}, {Kami{\'n}ski},
  {Udalski}, {Soszy{\'n}ski}, {Szyma{\'n}ski}, {Kubiak}, {Pietrzy{\'n}ski},
  {Poleski}, {Wyrzykowski}, \& {Ulaczyk}}]{Tylenda2011}
{Tylenda}, R., {Hajduk}, M., {Kami{\'n}ski}, T., {et~al.} 2011, \aap, 528, A114

\bibitem[{{Tylenda} \& {Kami{\'n}ski}(2016)}]{Tylenda2016}
{Tylenda}, R. \& {Kami{\'n}ski}, T. 2016, \aap, 592, A134

\bibitem[{{Tylenda} {et~al.}(2013){Tylenda}, {Kami{\'n}ski}, {Udalski},
  {Soszy{\'n}ski}, {Poleski}, {Szyma{\'n}ski}, {Kubiak}, {Pietrzy{\'n}ski},
  {Koz{\l}owski}, {Pietrukowicz}, {Ulaczyk}, \& {Wyrzykowski}}]{Tylenda2013}
{Tylenda}, R., {Kami{\'n}ski}, T., {Udalski}, A., {et~al.} 2013, \aap, 555, A16

\bibitem[{{Wesson} {et~al.}(2010){Wesson}, {Barlow}, {Ercolano}, {Andrews},
  {Clayton}, {Fabbri}, {Gallagher}, {Meixner}, {Sugerman}, {Welch}, \&
  {Stock}}]{Wesson2010}
{Wesson}, R., {Barlow}, M.~J., {Ercolano}, B., {et~al.} 2010, \mnras, 403, 474

\bibitem[{{Wilson} \& {Nordhaus}(2019)}]{Wilson2019}
{Wilson}, E.~C. \& {Nordhaus}, J. 2019, \mnras, 485, 4492

\end{thebibliography}
\bsp


\label{lastpage}
\end{document}